\documentclass[%
reprint,nofootinbib,
 amsmath,amssymb,
 aps,prd
floatfix,
]{revtex4-2}

\usepackage{graphicx}
\usepackage{dcolumn}
\usepackage{bm}

\usepackage{slashed}
\usepackage[utf8]{inputenc}
\usepackage{braket}

\newcommand{\sint}[2]{J_{#1}^{#2}}

\newcommand{\ppp}{p \cdot p^\prime }
\newcommand{\pps}{p^{\prime 2}}

\usepackage[compat=1.0.0]{tikz-feynman}

\usepackage[none]{hyphenat}

\begin{document}

\title{One-loop fermion-photon vertex in arbitrary gauge and dimensions: a novel approach }

\author{Victor Miguel Banda Guzm\'an}
\email{victor.banda@umich.mx}
\author{Adnan Bashir}
\email{adnan.bashir@umich.mx;abashir@jlab.org}

\affiliation{Instituto de F\'isica y Matem\'aticas
Universidad Michoacana de San Nicol\'as de Hidalgo
Edificio C-3, Apdo. Postal 2-82
C.P. 58040, Morelia, Michoac\'an, M\'exico}
\affiliation{Theory Center, Jefferson Lab, Newport News, VA 23606, USA}

\begin{abstract}

We compute one-loop electron-photon vertex with fully off-shell external momenta in an arbitrary covariant gauge and space-time dimension. There exist numerous efforts in literature where one-loop off-shell vertex is calculated by employing the standard first order Feynman rules in different covariant gauges and space-time dimensions of interest. The tensor structure which decomposes this three-point vertex into the components transverse and longitudinal to the photon momentum gets intertwined in this first order formalism. The Ward-Takahashi identity is explicitly invoked to untangle the two pieces and the results are expressed in a preferred basis of twelve spin-amplitudes.
We propose a novel approach based upon an
efficient combination of the first and second order formalisms of quantum electrodynamics to compute this one-loop vertex. Among some conspicuous advantages is the fact that this less known second order formalism separates the spin and scalar degrees of freedom of an electron interacting electromagnetically. More noticeably, the longitudinal and transverse contributions 
naturally disentangle from the onset in our approach. 
Moreover,  
this decomposition leads to identities between one-loop scalar Feynman integrals with higher powers in the propagators and shifted space-time dimensions that can be used to prove the Ward-Takahashi identity at one-loop order without the need to evaluate any Feynman integral. Additionally, this natural decomposition allows us to  establish the gauge-independence of the Pauli form factor through explicit cancellations of scalar Feynman integrals that depend on the gauge parameter. These cancellations naturally lead to a compact expression for the Pauli form factor in arbitrary dimensions.
Wherever necessary and insightful, we make comparisons with earlier works.
\end{abstract}

\maketitle

\section{Introduction}

Solving any quantum field theory (QFT) is equivalent to computing its Green functions. Three and four-point interaction vertices are the defining Green functions of any QFT within the standard model of particle physics. These vertices appear at the level of the Lagrangian. The simplest three-point vertex in quantum electrodynamics (QED) involves a photon interacting with a fermion, a charged lepton (like electron) or a quark. Therefore, the fermion-photon vertex not only orchestrates the dance of events when purely electromagnetic interactions are involved but it also plays a crucial role in unravelling the internal structure of hadrons. Though this structure and dynamics of hadrons are predominantly determined by quantum chromodynamics (QCD), it is generally probed through  electromagnetic interactions of photons with electrically charged quarks which compose all hadrons. One such example is provided by hadron electromagnetic form factors. The quark-photon vertex not only ensures charge conservation at zero photon momentum-transfer through the vector Ward-Takahashi identity (WTI) but its labyrinthine details also explain how the asymptotic limit of these form factors is faithfully approached for large momentum-transfer of the probing photons.   

A popular approach to study non-perturbative effects in a QFT in continuum is through the infinite set of Schwinger-Dyson equations (SDEs) which constitute the defining equations of motion of a QFT. Their intricate mathematical structure is such that the SDE of the $2$-point Green function is coupled to that of the $3$-point Green function through an integral equation, the one of the 3-point function is in turn entangled with that of the 4-point function {\em ad infinitum}. Any practical attempt to look for a non-perturbative solution is generally made by proposing a reliable {\em ansatz} for the relevant interaction vertices to truncate the infinite tower of these coupled equations to a finite solvable number.  It is only natural to assume that any non-perturbative construction, say that of the fermion-photon vertex, must agree with its perturbative expansion in the weak coupling regime. Thus the perturbative knowledge of this vertex can be used as a guiding tool for constructing such an {\em ansatz}.  
In literature, several such studies can be found, spanned over the last four decades. For example,
 \begin{itemize}
     \item  In~\cite{Ball_chiu}, one-loop QED vertex was computed in the Feynman gauge and four space-time dimensions, i.e., $D=4$. A convenient choice of the basis vectors was adopted therein such that the corresponding coefficients are free of  kinematic singularities.
     
     \item In Ref.~\cite{Curtis:1990zs}, one-loop vertex was reported in the asymptotic limit of momenta where momentum squared in one of the fermion legs is much greater than in the other. 
     
     \item The work of Ball and Chiu in Ref.~\cite{Ball_chiu} was extended to an arbitrary covariant gauge in Ref.~\cite{vertex_any_gauge}. The appearance of unwanted kinematic singularities for this general choice of an arbitrary covariant gauge was observed. However, this shortcoming was easily cured 
     by redefining two of the basis vectors as a superposition of the previous set of these vectors.  
     \item  In three space-time dimensions, $D=3$, this vertex for massless and massive fermions in an arbitrary covariant gauge was calculated in Refs.~\cite{vertex_3d_massless} and~\cite{vertex_3d}, respectively.
     \item
      For scalar QED, the one-loop vertex in arbitrary covariant gauge and dimensions was reported in Ref.~\cite{scalar_vertex}.  
     \item 
     A complete calculation of the quark-gluon vertex in an arbitrary covariant gauge and dimensions was carried out in detail in~\cite{vertex_QED_QCD}. All spinor QED results can be derived from the general expressions present therein through an appropriate choice of the color factors and selecting $D$ as desired. 
     \item  In Ref.~\cite{Numerical_vertex}, numerical and analytical expressions for the transverse and longitudinal form factors of the quark-gluon vertex in different kinematical regimes are presented. Again, one can deduce QED results from there.  
 \end{itemize}
 Following Ref.~\cite{Ball_chiu}, subsequent works commence with the WTI which relates the full fermion-photon vertex $\Gamma^{\mu}(p,p',k)$ with the full fermion propagator $S(p)$ not only at every order of perturbation theory but even more generally, non-perturbatively: 
\begin{eqnarray}
       k \cdot \Gamma(p,p',k) = S^{-1}(p')-S^{-1}(p) \;,
\end{eqnarray}
where $k$ is the incoming momentum of the photon, $p$ is that of the fermion and $p'$ is the outgoing momentum of the latter. This identity readily allows us to decompose the vertex into two components, the so called {\em longitudinal} piece $\Gamma^{\mu}_L(p,p',k)$ and the remaining part $\Gamma^{\mu}_T(p,p',k)$ which is transverse to the photon 4-momentum $k^{\mu}$~\footnote{Our metric in the Minkowski space is $\eta^{\mu\nu} = \text{diag}(-,+,+,+)$, while the Dirac gamma matrices satisfy the anticommutation relation $\{\gamma^\mu, \gamma^\nu \} = - 2 \eta^{\mu\nu}$.}:
\begin{eqnarray}
       \Gamma^{\mu}(p,p',k) = \Gamma^{\mu}_L(p,p',k) + \Gamma^{\mu}_T(p,p',k) \;,
\end{eqnarray}
where $\Gamma^{\mu}_T(p,p',k)$ satisfies the following restrictions:
\begin{eqnarray}
  k \cdot \Gamma_T(p,p',k) = 0 \;, \quad  \Gamma_T(p,p,0) = 0 \;.
\end{eqnarray}
The longitudinal part $\Gamma^{\mu}_L(p,p',k)$ alone satisfies the WTI. Starting from the limiting form of the WTI, namely, the Ward identity
\begin{eqnarray}
   \Gamma^{\mu}(p,p,0) = \frac{\partial}{\partial p_{\mu}} S^{-1}(p) \;,
\end{eqnarray}
Ball and Chiu proposed how to construct the longitudinal component, see~\cite{Ball_chiu} for the details and the explicit form of the vertex. We choose to call it the Ball-Chiu vertex and adopt the following notation
\begin{eqnarray}
       \Gamma^{\mu}_L(p,p',k) = \Gamma^{\mu}_{BC}(p,p',k) \;.
\end{eqnarray}
We would like to emphasize that neither the method of its construction nor the final form of the vertex is unique. However, it has become a standard practice to decompose the vertex in this manner:
\begin{eqnarray}
       \Gamma^{\mu}(p,p',k) = \Gamma^{\mu}_{BC}(p,p',k) + \Gamma^{\mu}_T(p,p',k) \;.
\end{eqnarray}
All one-loop results for the vertex are then presented by ensuing the following strategy:
\begin{itemize}
 \item The vertex in a particular kinematic configuration, gauge and dimensions is calculated at the one-loop level.
 \item 
 Accordingly, one-loop fermion propagator is also computed within the same set of requirements. 
 \item Using the last result,  $\Gamma^{\mu}_{BC}(p,p',k)$ is evaluated at the one-loop level: $\Gamma^{\mu \, 1-{\rm loop}}_{BC}(p,p',k)$. 
 \item One-loop transverse vertex is obtained by simply subtracting the longitudinal vertex from the full vertex:
\begin{eqnarray}
 \Gamma^{\mu \, 1-{\rm loop}}_{T}= \Gamma^{\mu \, 1-{\rm loop}} - \Gamma^{\mu \, 1-{\rm loop}}_{BC} \;. \nonumber
\end{eqnarray}
 \item This result is finally projected onto the transverse basis proposed in~\cite{Ball_chiu} or~\cite{vertex_any_gauge} and the transverse form factors are identified for each basis vector. 
\end{itemize}
The procedure outlined above is cumbersome though straightforward. There is a plethora of available research dedicated to constructing increasingly refined, reliable and physically meaningful {\em ans$\ddot{a}$tze}, taking perturbation theory as a guide where key elements of a  QFT, such as gauge invariance and renormalization, are satisfied order by order in a systematic manner. A comprehensive list of articles is hard to be provided without intentionally overlooking some of the several efforts in the literature. However, a selected list of references akin to the work we present here is~\cite{Curtis:1993py,Burden:1993gy,Bashir:1994az,Bashir:2000rv,Bashir:2004yt,Kizilersu:2009kg,Bashir:2011vg,Bashir:2011dp,Kizilersu:2014ela,Albino:2018ncl,Albino:2022efn}.

All the one-loop results we have discussed so far are obtained solely through employing the standard first order formalism of QED. In this article, we weave this approach with a relatively less scrutinized second order formalism, Refs.~\cite{Hostler,Morgan,WL_fprop1,WL_fprop2}. Explicit calculations presented here convincingly reveal that an adequate merger of both the first and the second order formalisms is particularly advantageous and efficient in the following aspects: \\

\noindent
{\bf{1.~Vertex decomposition:}}
  We shall observe in the next section that the electron-photon vertex $V^\mu(p,p^\prime, k)$ decomposes naturally into its longitudinal and transverse pieces at one-loop order without requiring a Ball-Chiu like decomposition\footnote{From now on, we shall adopt the notation $V^\mu$ for the one-loop vertex instead of $\Gamma^\mu$ which we will reserve for the full vertex.}: 
\begin{eqnarray}
V^\mu(p,p^\prime, k) = V_L^\mu(p,p^\prime, k) + V_T^\mu(p,p^\prime, k) \;, \label{vertex_dec}
\end{eqnarray} 
where $V_T^\mu$ satisfies the transversality condition
\begin{eqnarray}
k \cdot V_T(p,p^\prime, k) = 0 \;,
\end{eqnarray}
while the {\em longitudinal} piece $V_L^\mu$ satisfies the WTI at the same level of approximation:  
\begin{eqnarray}
k \cdot V_L(p,p^\prime, k) = e \left[ \Sigma(\slashed{p}) - \Sigma(\slashed{p}^\prime) \right], \label{WT_iden}
\end{eqnarray}
where $\Sigma(\slashed{p})$ is the fermion self-energy and $e$ is the usual QED coupling. We believe that this  feature of a natural decomposition of the full vertex into its longitudinal and transverse components would continue to persist at higher orders of the electromagnetic coupling. However, only an explicit calculation will be able to confirm this statement. \\

\noindent
{\bf{2.~Expeditious computation:}}
By employing an astute concoction of the first and the second order formalisms, we are able to select efficient routes to not only expedite the complete computation of one-loop vertex in arbitrary gauge and dimensions but also write it in a much more compact and concise form. However, in order to make a swift comparison with other results in the literature, we can always project our result onto the Ball-Chiu~\cite{Ball_chiu} or K{\i}zelers{${\rm \ddot{u}}$}-Pennington~\cite{vertex_any_gauge} basis. 


An added advantage of using this combined analysis is to keep track of those terms  which identically vanish when external momenta go on-shell at all intermediate stages of this calculation.
Both in the longitudinal and the transverse components of the vertex the terms which are explicitly irrelevant on-shell are ordered conveniently, e.g., the operator $(\slashed{p}+m)$ remains on the far right whereas $(\slashed{p}-m)$ is present on the left. Therefore, when on-shell conditions $(\slashed{p}+m)u_s(p)=0$ and $\overline{u}_s(p)(\slashed{p}+m)=0$ are imposed, they vanish. After eliminating these terms and using  the on-shell  symmetries of the Feynman integrals, we explicitly see the instantaneous cancellations of the integrals that depend explicitly on the covariant gauge parameter $\xi$, e.g., in the evaluation of the on-shell Pauli form factor. These cancellations naturally lead to a compact expression for it in an arbitrary space-time dimension in terms of scalar integrals with: (i) higher powers of scalar propagators and (ii) shifted dimensions. It hints towards a possible analogous simplification in the evaluation of the anomalous magnetic moment of the charged fermion at higher orders of perturbation theory.


The article is organized as follows: Sec.~\ref{summary_so} contains a summary of the second order formalism. In Sec.~\ref{vertex_combination}, we apply a convenient combination of first and second order formalisms to the fermion-photon vertex to decompose it into its longitudinal and transverse components. In Sec.~\ref{Longitudinal}, we provide explicit confirmation of the fact that the longitudinal part of the vertex is indeed {\em longitudinal}, i.e., it satisfies the WTI. We express it in terms of Feynman integrals. Then in 
Sec.~\ref{Transverse}, transverse contributions of the vertex which were previously obtained in Sec.~\ref{vertex_combination}, are written in terms of Feynman scalar integrals. Employing these expressions, and the results obtained in Sec.~\ref{Longitudinal} for the longitudinal part of the vertex, we also provide the coefficients of the Ball-Chiu basis in terms of scalar integrals. In Sec.~\ref{Form_factor} we calculate the Pauli form factor in $D$ dimensions, explicitly showing its independence of the covariant gauge parameter $\xi$ by cancellations of the scalar integrals that depend on this parameter in the longitudinal and transverse contributions of the vertex. A compact expression for the Pauli form factor in terms of scalar integrals is obtained. Making use of this general expression, special cases, i.e., $D=4$ and $D=3$ are reviewed, recovering the known results in the literature. Concluding remarks with prospects for future endeavors are provided in Sec.~\ref{conclusions}. The manuscript is complemented with three appendices with supplemental information. Appendix A provides explicit difference between the Ball-Chiu and our construction of the longitudinal vertex in perturbation theory. Naturally one is related to the other through the addition of a transverse piece.  Appendices B and C detail expressions of the transverse and longitudinal vertices, respectively, at one-loop order, and the scalar integrals involved with relevant identities relating them.   


\section{Second order formalism for QED} \label{summary_so}

\begin{figure}[h!]
\begin{center}
 \vspace{-1cm}
\includegraphics[scale=1]{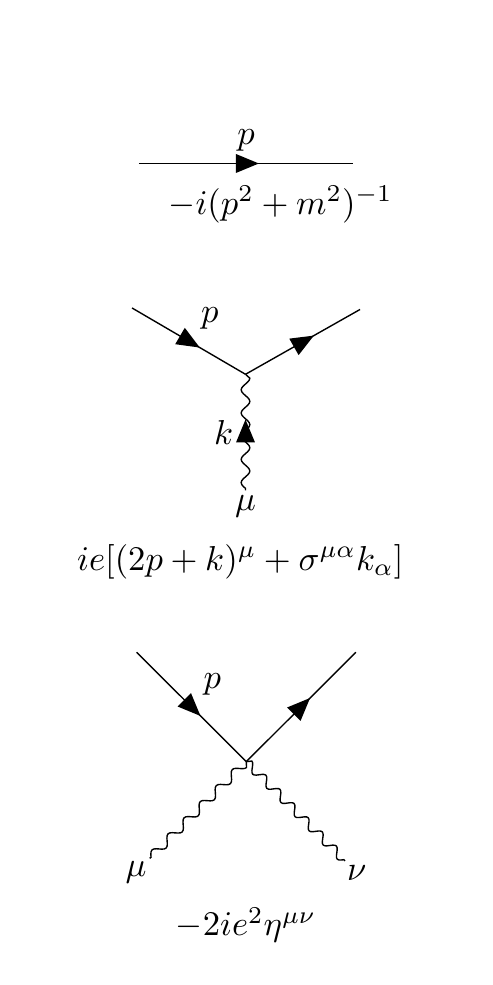}
\end{center}
\vspace{-1cm}
\caption{Second order rules for spinor QED} \label{SecondOR}
\end{figure}

The second order formalism for QED is based on Feynman rules displayed in Fig.~\ref{SecondOR} which were formally derived in~\cite{Morgan}. It provides an efficacious correspondence between spinor and scalar QED. This formalism has been used to compute fermion loops and amplitudes involving external fermions, see~\cite{Morgan} and references therein.   
In this article, we apply this strategy to achieve similar success in the off-shell evaluation of the QED vertex. 
The only difference is the additional spin factor $\sigma^{\mu\nu} = \frac{1}{2}[\gamma^\mu, \gamma^\nu]$ that appears in the QED vertex. It disentangles explicit contributions of the electromagnetic interaction of the fermionic particle into scalar and spin degrees of freedom with appropriate usage of Gordon-like identities. 
In addition to the rules of Fig.~\ref{SecondOR}, whenever a fermionic loop appears in a Feynman diagram, an extra factor of $-1/2$ must be included in the amplitude. 
Though the structure of the first order and the second order rules are different, the computed amplitude of a QED process following both formalisms is equivalent.

As mentioned before, one can derive the second order rules from the first order~\cite{Morgan}, by formally rewriting the product of the fermion propagator $S(p+k)$ with the first order vertex $e \gamma^\mu$ as,  
\begin{eqnarray}
e\, S(p+k) \, \gamma^\mu &=& 
e\,\dfrac{-(\slashed{p}+\slashed{k})+m }{(p+k)^2+m^2}\,  \gamma^\mu =
 \dfrac{A^{\mu}_{p,k}}{D_{p+k}}, \label{sor_dec}
\end{eqnarray}
where
\begin{eqnarray}
A^{\mu}_{p,k} &=& B^\mu_{p,k} + C_p^\mu, \nonumber \\
B^{\mu}_{p,k} &=& e(2p+k)^\mu + e\sigma^{\mu\alpha}k_\alpha, \nonumber \\
C_{p}^\mu &=& e \gamma^\mu (\slashed{p}+m), \nonumber \\
D_{q} &=& q^2 + m^2.  \label{ABC_def}
\end{eqnarray}
Here, $D_q^{-1}$ represents the scalar propagator, and $B_{p,k}$ is the three point vertex of the second order rules which naturally separates the scalar and spin electromagnetic interactions. Thus a product of a propagator and a vertex in a first order Feynman diagram can be decomposed into a second order contribution, and a left over first order term given by the operator $C_{p}^\mu$.

\begin{figure}[h!]
\begin{center}
 \vspace{-1cm}
\includegraphics[scale=1]{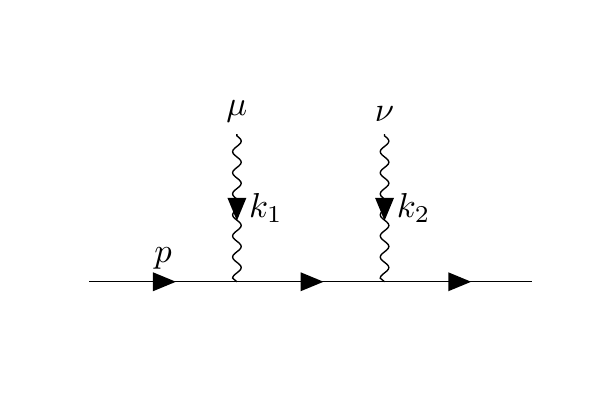}
\end{center}
\vspace{-1cm}
\caption{Subdiagram that produces the contribution to the four-point scalar vertex of the second order formalism.} \label{2Pairs_PV}
\end{figure}

Now, the four-point scalar vertex that appears in the second order rules arises when there are two consecutive pairs of a fermion propagator and a first order vertex in a Feynman diagram, as shown in Fig.~\ref{2Pairs_PV}. An amplitude constructed from a first order Feynman diagram that contains a subdiagram as the one shown in Fig.~\ref{2Pairs_PV} is proportional to $e^2 S(p+k_1+k_2) \gamma^\nu S(p+k_1) \gamma^\mu$. According to Eq.~\eqref{sor_dec}, it can be decomposed as,
\begin{eqnarray}
\hspace{-2mm} e^2 S(p+k_1+k_2) \gamma^\nu S(p+k_1) \gamma^\mu \hspace{-1mm} = \hspace{-1mm} \dfrac{A^\nu_{p+k_1, k_2}}{D_{p+k_1+k_2}} \dfrac{A^\mu_{p+k_1}}{D_{p+k_1}}.  \label{4Point_der} 
\end{eqnarray}
Since
\begin{eqnarray}
C_{p+k}^\mu \; \dfrac{A_{p,k}^\nu}{D_{p+k}} &=& e^2 \gamma^\mu \gamma^\nu 
  - e^2 \left( \eta^{\mu\nu} - \sigma^{\mu\nu} \right), \label{C_prop}
\end{eqnarray}
Eq.~\eqref{4Point_der} acquires the following form
\begin{eqnarray}
&& e^2 S(p+k_1+k_2) \gamma^\nu S(p+k_1) \gamma^\mu = \nonumber \\
&& \hspace*{4em} \dfrac{B^\nu_{p+k_1, k_2}}{D_{p+k_1+k_2}} \dfrac{A^\mu_{p+k_1}}{D_{p+k_1}}
- e^2 \dfrac{\eta^{\mu\nu}+\sigma^{\mu\nu}}{D_{p+k_1+k_2}}.  \label{4point_der2}
\end{eqnarray}

Thus, the multiplication of fermion propagators and first order vertices on the left-hand side of this equation generates a term that is proportional to $\eta^{\mu\nu}$ and has one less power of the scalar propagator. It readily gets identified as a four-point scalar vertex. 

Following the procedure where fermion propagators and first order vertices are written according to Eq.~\eqref{sor_dec} and  applying identity~\eqref{C_prop}, any amplitude constructed from the first order rules can be obtained equivalently and conveniently from the second order rules given in Fig.~\eqref{SecondOR}. The proof  of this equivalence can be found in Ref.~\cite{Morgan}. 
We now proceed to show how this systematic procedure 
can generate a natural decomposition of the QED vertex into its longitudinal and transverse pieces.

\section{The QED vertex at one-loop from the combination of the first and second order formalisms} \label{vertex_combination}

\begin{figure}[h!]
\begin{center}
\vspace{-1cm}
\includegraphics[scale=1]{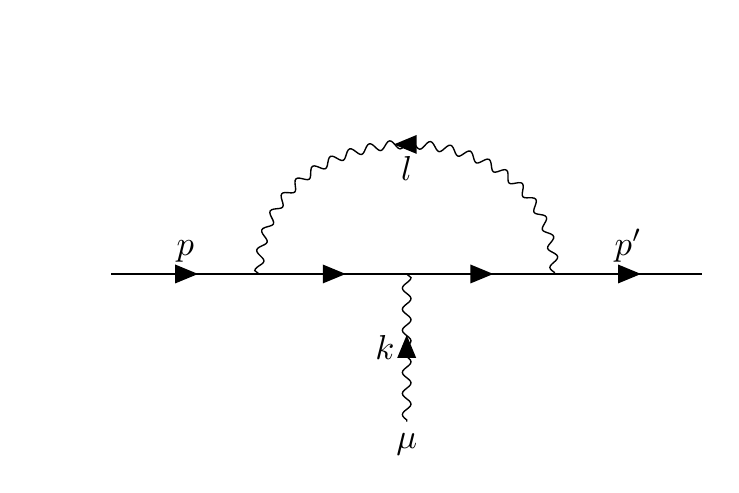}
\vspace{-1cm}
\end{center}
\caption{One-loop vertex $V^\mu(p',p)$ in the first order formalism.} \label{1LoopD}
\end{figure}

Fig.~\ref{1LoopD} depicts the Feynman diagram for the vertex function $V^\mu$ at one-loop within the first order formalism. Following the standard Feynman rules, its  mathematical expression is:
\begin{equation}
V^\mu(p^\prime, p) = \dfrac{e^3}{i} \int \dfrac{d^Dl}{(2\pi)^D} \left[ \gamma^\rho S(p^\prime+l)\gamma^\mu S(p+l) \gamma^\nu \right] \Delta_{\nu\rho}(l), \label{Vert_1L} 
\end{equation} 
where the photon propagator $\Delta_{\mu\nu}(l)$ in an arbitrary covariant gauge $\xi$ is given by:
\begin{equation}
\Delta_{\mu\nu}(l) = \dfrac{1}{l^2}\left( \eta_{\mu\nu}-(1-\xi)\dfrac{l_\mu l_\nu}{l^2} \right).
\end{equation}
Using Eq.~\eqref{sor_dec}, the vertex $V^\mu$ of Eq.~\eqref{Vert_1L} becomes:
\begin{eqnarray}
\hspace{-2mm} V^\mu(p^\prime, p) = e \hspace{-1mm} \int \hspace{-1mm} \dfrac{d^Dl}{i(2\pi)^D}  \gamma^\rho \left( \dfrac{A_{p+l,k}^\mu}{D_{p^\prime+l}} \right) \hspace{-1mm} \left( \dfrac{A^\nu_{p,l}}{D_{p+l}} \right) \hspace{-1mm} \Delta_{\rho\nu}(l), 
\end{eqnarray}

which can be written as:
\begin{eqnarray}
V^\mu(p^\prime, p) &=& e\, \int \dfrac{d^Dl}{i(2\pi)^D} \, \gamma^\rho \Bigg( \dfrac{ B^\mu_{p+l,k}\, B^\nu_{p,l}}{D_{p^\prime+l} D_{p+l}} + \dfrac{ B^\mu_{p+l,k}\, C^\nu_{p}}{D_{p^\prime+l} D_{p+l}} \nonumber \\
&& + e^2 \dfrac{\gamma^\mu\, \gamma^\nu}{D_{p^\prime+l}} \Bigg) \Delta_{\rho\nu}(l), \label{V_12_com}
\end{eqnarray}
after application of identity \eqref{C_prop}, and the definitions given in Eqs.~\eqref{ABC_def}. Here $V^\mu$ is expressed as a combination of first and second order vertices, with the corresponding scalar propagators of the Feynman diagram of Fig.~(\ref{1LoopD}).
The numerator of the third term on the right-hand side of Eq.~\eqref{V_12_com} can be rearranged as
\begin{eqnarray}
\hspace{-2mm}
 {l^4} \gamma^\rho \, \gamma^\mu\, \gamma^\nu \,   \Delta_{\rho\nu}(l) = { l^2 (D-3+\xi) \gamma^\mu + 2  (1-\xi)l^\mu\slashed{l}}, \label{V3}
\end{eqnarray}
where we have used the following identity:
\begin{eqnarray}
 {l^2} \gamma^\rho \, \gamma^\nu \, \Delta_{\rho\nu} = {{1-D-\xi}}. \label{gg_pp_iden}
\end{eqnarray}
The result that stems from expression~\eqref{V3} will form part of the longitudinal  vertex as we show later. On the other hand, the second term on the right-hand side of Eq.~\eqref{V_12_com} can be split into longitudinal and  transverse pieces by commuting $\gamma^\rho$ with $B^\mu_{p+l,k}$. Using the identity
\begin{eqnarray}
\gamma^\mu\, B^\nu_{p,k} = B^\nu_{p,k}\, \gamma^\mu - 2 e\, \eta^{\mu\nu} \slashed{k} + 2e\, k^\mu \gamma^\nu, \label{comm_gB}
\end{eqnarray}
the numerator of the second term acquires the form:
\begin{eqnarray}
&& \gamma^\rho \, B^\mu_{p+l,k}\, C^\nu_{p} \, \Delta_{\rho\nu} \nonumber \\
&& = e  \Big( B^\mu_{p+l,k}\, \gamma^\rho \,\gamma^\nu - 2 e\, \eta^{\rho\mu} \slashed{k} \,\gamma^\nu + 2e\, k^\rho\, \gamma^\mu \,\gamma^\nu \Big) \nonumber \\
&& \hspace*{8pt} \times \Delta_{\rho\nu} \left( \slashed{p}+m \right) \nonumber \\ 
&& = \dfrac{e^2}{ l^2} \Big[ (1-D-\xi)(p+p^\prime+2l)^\mu +  \dfrac{2(1-\xi)}{l^2}\Big(l^\mu \slashed{k} \slashed{l} \nonumber \\
&& \hspace*{8pt} - k \cdot l\, \gamma^\mu \slashed{l} \Big) + (5-D-\xi) \sigma^{\mu\alpha}k_\alpha \Big] (\slashed{p}+m), \label{V2} 
\end{eqnarray}
where we have made use of identity~\eqref{gg_pp_iden}. It can be readily observed that only the term proportional to $(2p+k+2l)^\mu$ pertains to the longitudinal vertex; the other terms are transverse.
Similar procedure allows us to re-write the numerator of the first term on the right-hand side of Eq.~\eqref{V_12_com}
as follows:
\begin{eqnarray}
 \gamma^\rho \, B^\mu_{p+l,k}\, B^\nu_{p,l} \, \Delta_{\rho\nu} &=& B^\mu_{p+l,k}\, \gamma^\rho\, B^\nu_{p,l}\, \Delta_{\rho\nu} - 2 e\, \slashed{k}\, B^\nu_{p,l}\,\Delta_{\mu\nu} \nonumber \\
&&  + 2 e\, k^\rho \, \gamma^\mu \, B^\nu_{p,l}\,\Delta_{\rho\nu}.  \label{V1_1}
\end{eqnarray}
Since
\begin{eqnarray}
\gamma^\rho\, B^\nu_{p,l}\, \Delta_{\rho\nu} &=& \dfrac{e}{l^2}\Bigg[ 2 \slashed{p} + \Bigg( 1-D+\xi \nonumber \\
&& -(1-\xi)\dfrac{2 p\cdot l}{l^2} \Bigg) \slashed{l} \Bigg], \nonumber \\
\slashed{k} \, B^\nu_{p,l}\, \Delta_{\mu\nu} &=& \dfrac{1}{l^2} \Big[ B^\mu_{p,l}\,\slashed{k} + 2e(k\cdot l \gamma^\mu-k^\mu \slashed{l}) \nonumber \\
&& -e(1-\xi) \dfrac{2 p \cdot l + l^2}{l^2} l^\mu \, \slashed{k} \Big], \nonumber \\
k^\rho \, B^\nu_{p,l}\,\Delta_{\rho\nu} &=& \dfrac{e}{l^2} \Bigg[ k \cdot (2p+l) + \sigma^{\beta\alpha} l_\alpha k_\beta \nonumber \\
&& -(1-\xi) k \cdot l \left( 2 \dfrac{ p \cdot l }{l^2} + 1 \right) \Bigg],
\end{eqnarray}
we can now make use of these detailed identities to cast Eq.~\eqref{V1_1} in the following form
\begin{eqnarray}
&& \gamma^\rho \, B^\mu_{p+l,k}\, B^\nu_{p,l} \, \Delta_{\rho\nu} = \nonumber \\
&& \dfrac{e^2}{l^2} \Big[ 2 \left[ (p+p^\prime +2 l)^\mu + \sigma^{\mu\alpha} k_\alpha \right] \slashed{p}  + 4(p\cdot k\, \gamma^\mu-p^\mu \slashed{k})
\nonumber \\
&& + 2 \xi \left( l \cdot k \, \gamma^\mu - l^\mu \, \slashed{k} \right) + 4 \left( \gamma^\mu \slashed{k} \slashed{l} + k^\mu \, \slashed{l} \right) \nonumber \\
&& + 2 \left( k\cdot l \,\gamma^\mu - l^\mu \, \slashed{k} \right) + (1-D+\xi) \big[ (p+p^\prime + 2 l)^\mu  \nonumber \\
&& + \sigma^{\mu\alpha}k_\alpha \big] \slashed{l}   - 2\,\dfrac{1-\xi}{l^2} \Big\{ p \cdot l \left[ (p+p^\prime + 2 l)^\mu  + \sigma^{\mu\alpha}k_\alpha \right] \slashed{l} \nonumber \\ 
&& + 2\left[ l \cdot k p \cdot l \gamma^\mu - p \cdot l l^\mu \, \slashed{k}\right] \Big\} \Big]. \label{V1}
\end{eqnarray}
Again we have an unambiguous separation between the longitudinal and  transverse components of the vertex. The former correspond to the terms proportional to $(p+p^\prime)\slashed{p}$, $l^\mu \slashed{p}$, $(p+p^\prime)\slashed{l}$ and $l^\mu \slashed{l}$, while the transverse parts form the remaining terms. According to Eqs.~\eqref{V3},~\eqref{V2},~\eqref{V1}, the longitudinal and transverse components, $V^\mu_L$ and $V^\mu_T$, respectively, defined in Eq.~\eqref{vertex_dec}, read as,

\begin{widetext}
\begin{eqnarray}
V^\mu_L &=& e^3 \int \dfrac{d^Dl}{i(2\pi)^D}\, \Bigg\{ \dfrac{1}{D_{p^\prime+l}D_{p+l}l^2}\Bigg[ 2(p+p^\prime + 2 l)^\mu \slashed{p} + (1-D+\xi)(p+p^\prime)^\mu \slashed{p} \nonumber + 2(1-D+\xi)l^\mu \slashed{l} \nonumber \\
&-& 2 (1-\xi)(p+p^\prime + 2 l)^\mu \dfrac{p \cdot l \slashed{l}}{l^2}  + (1-D-\xi)\left[ (p+p^\prime)^\mu + 2 l^\mu \right](\slashed{p}+m) \Bigg] + \dfrac{D-3+\xi}{D_{p^\prime+l} \, l^2} \gamma^\mu + 2 \dfrac{(1-\xi) l^\mu \, \slashed{l}}{D_{p^\prime+l} \, l^4} \Bigg\},  \label{VL_l} 
\end{eqnarray}
\begin{eqnarray}
V^\mu_T &=& e^3 \int \dfrac{d^Dl}{i(2\pi)^D}\, \dfrac{1}{D_{p^\prime+l}D_{p+l}l^2} \Bigg\{ \left[ (5-D-\xi)\sigma^{\mu\alpha}k_\alpha + \dfrac{2(1-\xi)}{l^2}\left( l^\mu \slashed{k} \slashed{l}- k \cdot l \gamma^\mu \slashed{l} \right) \right](\slashed{p}+m) + 2 \sigma^{\mu\alpha}k_\alpha \slashed{p} \nonumber \\
&+& 4 (p\cdot k \gamma^\mu -p^\mu \slashed{k} )
+ 2 \xi \left( k \cdot l \gamma^\mu - l^\mu \slashed{k} \right) +  (1-D+\xi)\sigma^{\mu\alpha}k_\alpha \slashed{l} \nonumber + 4 \left( \gamma^\mu \slashed{k} \slashed{l} + k^\mu \slashed{l} \right) + 2 (k \cdot l \gamma^\mu - l^\mu \slashed{k} ) \nonumber \\
&-& 2(1-\xi) \dfrac{p \cdot l}{l^2} \sigma^{\mu\alpha}k_\alpha \slashed{l} + 4 \dfrac{1-\xi}{l^2}\left( p \cdot l \slashed{k} l^\mu - l \cdot k p \cdot l \gamma^\mu  \right) \Bigg\}. \label{VT_l}
\end{eqnarray}
\end{widetext}

From the expression above for $V_T^\mu$, one can readily infer that it satisfies the transversality condition $k \cdot V_T = 0$. However, it is less obviously discernible that $V_L^\mu$ satisfies the WTI given in Eq.~\eqref{WT_iden}. However, we set out to show this explicitly in the next section.  

\section{$V_L^\mu$ and the Ward-Takashi identity} \label{Longitudinal}

In order to demonstrate that the longitudinal vertex $V_L^\mu$ given by Eq.~\eqref{VL_l} indeed satisfies the WTI at the one-loop level, a natural starting point is to contract it with the photon momentum $k_\mu$:
\begin{eqnarray}
k \cdot V_L &=& e^3 \int \dfrac{d^Dl}{i(2\pi)^D}\,\Bigg\{ 
\dfrac{1}{D_{p^\prime+l}D_{p+l}\,l^2} \Big[ (1-D-\xi) \nonumber \\
&& \times [ k \cdot (p+p^\prime) + 2 k\cdot l ] (m+\slashed{l}) + (3-D-\xi) \nonumber \\
&& \times [k \cdot (p+p^\prime) + 2 k\cdot l ] \, \slashed{p}  \Big] + \dfrac{D-3+\xi}{D_{p^\prime+l}\, l^2}\, \slashed{k} \nonumber \\
&& + \dfrac{1-\xi}{D_{p^\prime+l}D_{p+l}\,l^4} \Big[ 2 (p^2+m^2) p^\prime \cdot l + 2 l \cdot p^\prime l^2 \nonumber \\ 
&&  -(p^{\prime 2}+m^2) p \cdot l - 2 l \cdot p l^2 \Big] \slashed{l} \Bigg\}.  \label{kV_1} 
\end{eqnarray}
Perhaps the most efficient way to proceed is to reduce the vector and tensor integrals involved into scalar integrals. We can then make use of the known identities connecting these scalar integrals to simplify the resulting expression and rearrange it in a convenient manner so that the final terms can readily be identified with the expressions which define electron self energy. One way to achieve this goal is to employ the tensor reduction algorithm described first in Ref.~\cite{Davydychev:1991va} and later also used in Refs.~\cite{Fiorentin}-\cite{Anastasiou3}. It allows us to write any multi-loop tensor integrals in terms of scalar integrals:
\begin{eqnarray}
\int \dfrac{d^D l }{i \pi^{D/2}} \dfrac{l^\mu}{D_{p^\prime + l} \, D_{p+l} \; l^{2 \, a} } &=& - p^\mu \; J_{1, 2, a}^{D+2} - p^{\prime\mu} \; J_{2, 1, a}^{D+2}, \nonumber \\
\int \dfrac{d^D l }{i \pi^{D/2}} \dfrac{l^\mu l^\nu}{ D_{p^\prime + l} \, D_{p+l}  \; l^{2 \, a} }
&=& \dfrac{1}{2} \eta^{\mu\nu} \; J_{1, 1, a}^{D+2} + 2 p^{\prime \mu} p^{\prime \nu}  \; \sint{3,1,a}{D+4} \nonumber \\  \nonumber \\ 
&& \hspace{-2.5cm} + 2 p^{\mu} p^{ \nu}  \; \sint{1,3,a}{D+4} + (p^\mu p^{\prime \nu} + p^\nu p^{\prime \mu} ) \; \sint{2,2,a}{D+4}, \label{integrals}
\end{eqnarray}
where
\begin{eqnarray}
J_{a, b, c}^{\tilde{D}}(p,p') = \int \dfrac{d^{\tilde{D}} l }{i \pi^{\tilde{D}/2}} \dfrac{1}{ D_{p^\prime + l}^a \, D_{p+l}^b  \; l^{2 \, c} }.   \label{J_def}
\end{eqnarray}
Using this tensor reduction algorithm multiple times, Eq.~\eqref{kV_1} can be rearranged in the following form:  
\begin{eqnarray}
&& k \cdot V_L = \nonumber \\ \nonumber \\
&& \dfrac{e^3}{(4\pi)^{\frac{D}{2}}} \Bigg\{ m(1-D-\xi) \Big[ (p^{\prime 2}-p^2)\sint{1,1,1}{D} - 2 k \cdot p \sint{1,2,1}{D+2} \nonumber \\ \nonumber \\
&& -2 k \cdot p^\prime \sint{2,1,1}{D+2} \Big] + \Big( (3-D-\xi)\Big[ (p^{\prime 2}-p^2)\sint{1,1,1}{D} \nonumber \\ \nonumber \\
&& - 2 k \cdot p \sint{1,2,1}{D+2}  - 2 k \cdot p^\prime \sint{2,1,1}{D+2} + \sint{1,0,1}{D} \Big] \nonumber \\\nonumber \\
&& +  (2-D)\Big[ (p^2-p^{\prime 2})\sint{1,2,1}{D+2} - \sint{1,1,1}{D+2} + 2 k \cdot p^\prime \sint{2,2,1}{D+4} \nonumber \\ \nonumber \\
&&  + 4 k \cdot p \sint{1,3,1}{D+4} \Big] + (1-\xi) \Big[ (p^{\prime 2}-p^2) \sint{1,2,1}{D+2} \nonumber \\ \nonumber \\
&& + (p^2+m^2) \left( 2 p^{\prime 2} \sint{2,2,2}{D+4} + 4 p \cdot p^\prime \sint{1,3,2}{D+4} \right) \nonumber \\\nonumber \\
&& - (p^{\prime 2}+m^2) \Big( \sint{1,1,2}{D+2} + 2 p \cdot p^\prime \sint{2,2,2}{D+4} + 4 p^2 \sint{1,3,2}{D+4} \Big) \, \Big] \; \Big) \; \slashed{p} \nonumber \\ \nonumber \\
&&   + \Bigg(  (2-D)\Big[ (p^2-p^{\prime 2}) \sint{2,1,1}{D+2} + \sint{1,1,1}{D+2} + 2 k \cdot p \sint{2,2,1}{D+4} \nonumber \\ \nonumber \\
&& + 4 k \cdot p^\prime \sint{3,1,1}{D+4} \Big] + (1-\xi) \Big[ (p^{\prime 2}-p^2) \sint{2,1,1}{D+2} \nonumber \\
\nonumber \\
&&  + (p^2+m^2)\left( \sint{1,1,2}{D+2} + 2 p \cdot p^\prime \sint{2,2,2}{D+4} + 4 p^{\prime 2} \sint{3,1,2}{D+4} \right) \nonumber \\ \nonumber \\
&& -(p^{\prime 2}+m^2)\left( 2 p^2 \sint{2,2,2}{D+4} + 4 p \cdot p^\prime \sint{3,1,2}{D+4} \right) \Big] \nonumber \\
&& + (D-3+\xi) \sint{1,0,1}{D}  \Bigg) \slashed{p}^\prime \Bigg\}. \label{kV_2}
\end{eqnarray} 
All the scalar integrals appearing in Eq.~\eqref{kV_2} can be expressed as a linear combination of the master integrals $\sint{1,1,1}{D}$, $\sint{0,1,0}{D}$, $\sint{0,1,1}{D}$, $\sint{1,0,1}{D}$ and $\sint{1,1,0}{D}$ by implementing the well-established and widely employed {\em Integration by Parts Technique} (IBP)~\cite{IBP_alg, IBP_letter} as well as {\em dimensional recurrence relations}~\cite{Tarasov_DRR, Lee_DRR1, Lee_DRR2}. These methods, aided with the symbolic programming package LiteRed~\cite{LiteRed1, LiteRed2}, yield the following practically useful identities:
\begin{eqnarray}
\sint{0,1,1}{D} - \sint{1,0,1}{D} &=& (p^{\prime 2}-p^2)\sint{1,1,1}{D} - 2 k \cdot p \sint{1,2,1}{D+2} \nonumber \\
&& -2 k \cdot p^\prime \sint{2,1,1}{D+2}, \label{iden1} \\
\sint{0,1,0}{D}+(m^2+p^2)\sint{0,1,1}{D} &=& -2p^2\Big[ (p^2-p^{\prime 2})\sint{1,2,1}{D+2} \nonumber \\
&& - \sint{1,1,1}{D+2} + 2 k \cdot p^\prime \sint{2,2,1}{D+4} \nonumber \\
&& + 4 k \cdot p \sint{1,3,1}{D+4} \Big],  \label{iden2}\\
\sint{0,1,0}{D} + (p^{\prime 2}+m^2)\sint{1,0,1}{D} &=& 2 p^{\prime 2} \Big[ 
(p^2-p^{\prime 2}) \sint{2,1,1}{D+2} \nonumber \\
&& + \sint{1,1,1}{D+2} + 2 k \cdot p \sint{2,2,1}{D+4}  \nonumber \\
&& + 4 k \cdot p^\prime \sint{3,1,1}{D+4} \Big],  \label{iden3}
\end{eqnarray}
and
\begin{eqnarray}
(2-D) \sint{0,1,0}{D} &=& \left[ (4-D)p^2 + (D-2) m^2 \right] \sint{0,1,1}{D}  \nonumber \\ 
&& + 2p^2 \Big[ (p^{\prime 2}-p^2) \sint{1,2,1}{D+2} + (p^2+m^2) \nonumber \\
&& \times \left( 2 p^{\prime 2} \sint{2,2,2}{D+4} + 4 p \cdot p^\prime \sint{1,3,2}{D+4} \right) \nonumber \\
&&  - (p^{\prime 2}+m^2) ( \sint{1,1,2}{D+2} + 2 p \cdot p^\prime \sint{2,2,2}{D+4} \nonumber \\
&& + 4 p^2 \sint{1,3,2}{D+4} ) \Big], \label{iden4} \\
(2-D) \sint{0,1,0}{D} &=& \left[ (4-D)p^{\prime\,2} + (D-2) m^2 \right] \sint{1,0,1}{D}  \nonumber \\ 
&& + 2 p^{\prime\,2}\Big[ (p^{\prime 2}-p^2) \sint{2,1,1}{D+2} -(p^{\prime 2}+m^2) \nonumber \\
&& \times \left( 2 p^2 \sint{2,2,2}{D+4} + 4 p \cdot p^\prime \sint{3,1,2}{D+4} \right) \nonumber \\
&& + (p^2+m^2) ( \sint{1,1,2}{D+2} + 2 p \cdot p^\prime \sint{2,2,2}{D+4} \nonumber \\
&& + 4 p^{\prime 2} \sint{3,1,2}{D+4} ) \Big]. \label{iden5}
\end{eqnarray}
Using these identities, Eq.~\eqref{kV_2} simplifies to
\begin{eqnarray}
\hspace{-1cm} k \cdot V_L &=& \dfrac{e^3}{(4\pi)^{\frac{D}{2}}} \Big\{ m (1-D-\xi) \left( \sint{0,1,1}{D} - \sint{1,0,1}{D} \right) \nonumber \\
&+&  \dfrac{D-2}{2 p^2} \xi \left[ \sint{0,1,0}{D} + (m^2-p^2) \sint{0,1,1}{D} \right] \slashed{p} \nonumber \\
&-& \dfrac{D-2}{2 p^{\prime 2}} \xi \left[ \sint{0,1,0}{D} + (m^2-p^{\prime 2}) \sint{1,0,1}{D} \right] \slashed{p}^\prime \Big\}. 
\end{eqnarray}
Since fermion self-energy $\Sigma(\slashed{p})$ at one-loop in an arbitrary covariant gauge $\xi$ and dimensions $D$ is given by \cite{WL_fprop1}:
\begin{eqnarray}
\Sigma(\slashed{p}) &=& \dfrac{e^2}{(4\pi)^{\frac{D}{2}}} \Big\{  m (1-D-\xi) \sint{0,1,1}{D} \nonumber \\
&+&  \dfrac{D-2}{2 p^2} \xi \left[ \sint{0,1,0}{D} + (m^2-p^2) \sint{0,1,1}{D} \right] \slashed{p} \Big\}, 
\end{eqnarray}
it can be readily inferred that $V_L^\mu$ satisfies the WTI, Eq.~\eqref{WT_iden}, as expected.

To wind up this section, let us write down the expression for $V_L^\mu$ in terms of Feynman scalar integrals. Using Eq.~\eqref{VL_l} and the tensor reduction algorithm, $V_L^\mu$ reads as:
\begin{eqnarray}
V_L^\mu &=& \dfrac{e^3}{(4\pi)^{\frac{D}{2}}} \Big\{ 
 (D-1-\xi)(p+p^\prime)^\mu \left( \sint{1,2,1}{D+2} \slashed{p} + \sint{2,1,1}{D+2} \slashed{p}^\prime \right) \nonumber \\
&& + 2 \sint{1,1,1}{D} (p+p^\prime)^{\mu} \slashed{p} - 4 \left( \sint{1,2,1}{D+2} p^\mu + \sint{2,1,1}{D+2} p^{\prime \mu}  \right) \slashed{p} \nonumber \\
&& + (2-D)\Big[ \sint{1,1,1}{D+2} \, \gamma^\mu + 2 \sint{2,2,1}{D+4} (p^{\prime \mu} \slashed{p} + p^\mu \slashed{p}^\prime)  \nonumber \\ 
&& + 4 \sint{1,3,1}{D+4} p^\mu \slashed{p} + 4 \sint{3,1,1}{D+4} p^{\prime \mu} \slashed{p}^\prime  \Big] \nonumber \\
&& + (1-D-\xi) \Big[ \sint{1,1,1}{D}(p+p^\prime)^\mu - 2 \sint{1,2,1}{D+2} p^\mu \nonumber \\ 
&& - 2 \sint{2,1,1}{D+2} p^{\prime \mu} \Big](\slashed{p}+m) + (D-3+\xi) \sint{1,0,1}{D} \, \gamma^\mu \nonumber \\
&& + (1-\xi)(p^2+m^2)\Big[ \sint{1,1,2}{D+2}\, \gamma^\mu + 4 \sint{1,3,2}{D+4}p^\mu \slashed{p} \nonumber \\
&& + 4 \sint{3,1,2}{D+4} p^{\prime \mu} \slashed{p}^\prime + 2 \sint{2,2,2}{D+4} (p^{\prime \mu} \slashed{p} + p^\mu \slashed{p}^\prime ) \Big]   \nonumber \\
&&    - (1-\xi)(p+p^\prime)^\mu \Big[ \sint{1,1,2}{D+2} \slashed{p} + 4 p \cdot p^\prime \sint{3,1,2}{D+4} \slashed{p}^\prime \nonumber \\
&& + 4 p^2 \sint{1,3,2}{D+4} \slashed{p} + 2 \sint{2,2,2}{D+4} ( p \cdot p^\prime \slashed{p} + p^2 \slashed{p}^\prime )  \Big] \Big\}.  \label{VL_1}
\end{eqnarray}
In the basis $\{ p^\mu, p^{\prime \mu}, \gamma^\mu, p^{\mu} \slashed{p}, p^{\mu} \slashed{p}^\prime, p^{\prime \mu} \slashed{p}, p^{\prime \mu} \slashed{p}^\prime \}$ it can be rewritten as follows:
\begin{eqnarray}
&& \hspace{-9mm} V_L^\mu = \dfrac{e^3}{(4\pi)^{\frac{D}{2}}} \Big\{ m(1-\xi-D)(\sint{1,1,1}{D}-2\sint{1,2,1}{D+2})p^\mu \nonumber \\
&& \hspace{-3.5mm} + m(1-\xi-D)(\sint{1,1,1}{D}-2\sint{2,1,1}{D+2})p^{\prime \mu} \nonumber \\
&& \hspace{-3.5mm}  + \Big[ (2-D) \sint{1,1,1}{D+2} + (D-3+\xi) \sint{1,0,1}{D} \nonumber \\
&& \hspace{-3.5mm}  + (m^2+p^2)(1-\xi)\sint{1,1,2}{D+2} \Big] \gamma^\mu \nonumber \\
&& \hspace{-3.5mm}  + \Big[ (2-D)\left( \sint{1,1,1}{D} - 3 \sint{1,2,1}{D+2} + 4 \sint{1,3,1}{D+4} \right) \nonumber \\
&& \hspace{-3.5mm}  + (1-\xi)\big( \sint{1,1,1}{D} - \sint{1,1,2}{D+2} - \sint{1,2,1}{D+2}  \nonumber \\
&& \hspace{-3.5mm}  + 4 m^2 \sint{1,3,2}{D+4} - 2 p \cdot p^\prime \sint{2,2,2}{D+4} \big) \Big] p^\mu \slashed{p} \nonumber \\
&& \hspace{-3.5mm}  + \Big[ (D-2)\left( \sint{2,1,1}{D+2} - 2 \sint{2,2,1}{D+4} \right) + (1-\xi)\big( \sint{2,1,1}{D+2} \nonumber \\ 
&& \hspace{-3.5mm}  + 2m^2 \sint{2,2,2}{D+4} - 4 p \cdot p^\prime \sint{3,1,2}{D+4} \big) \Big] p^\mu \slashed{p}^\prime \nonumber \\
&& \hspace{-3.5mm}  + \Big[ (2-D)\big( \sint{1,1,1}{D} - \sint{1,2,1}{D+2} - 2 \sint{2,1,1}{D+2} + 2 \sint{2,2,1}{D+4}\big) \nonumber \\ 
&& \hspace{-3.5mm}  + (1-\xi) \Big( \sint{1,1,1}{D} - \sint{1,1,2}{D+2} + \sint{1,2,1}{D+2} - 4 p^2 \sint{1,3,2}{D+4}  \nonumber \\
&& \hspace{-3.5mm}  - 2 \sint{2,1,1}{D+2} + 2 (m^2+p^2-p \cdot p^\prime) \sint{2,2,2}{D+4} \big) \Big] p^{\prime \mu } \slashed{p} \nonumber \\
&& \hspace{-3.5mm} + \Big[ (D-2) \left( \sint{2,1,1}{D+2}-4\sint{3,1,1}{D+4} \right) + (1-\xi)\Big( \sint{2,1,1}{D+2} \nonumber \\
&& \hspace{-3.5mm} -2p^2 \sint{2,2,2}{D+4} + 4(m^2+p^2-p\cdot p^\prime)\sint{3,1,2}{D+4} \Big) \Big] p^{\prime \mu} \slashed{p}^\prime \Big\}.  \label{VL_2}
\end{eqnarray}
The scalar integrals sprinkled all over Eq.~\eqref{VL_2} contain higher powers of propagators and have shifted space-time dimensions. These are explicitly detailed in appendix~\ref{s_int} as a linear combination of elementary master integrals whose solutions are well-known in literature in various space-time dimensions $D$. We can now proceed to carry out a similar analysis for the transverse vertex in the next section. As the details have already been outlined, we will merely present the results. 

\section{$V_T^\mu$ and the Ball-Chiu Basis} \label{Transverse}

One-loop transverse vertex $V_T^\mu$ is given by Eq.~\eqref{VT_l}. On applying tensor reduction algorithm it can be written in terms of scalar Feynman integrals as follows:
\begin{eqnarray}
&& V_T^\mu = \nonumber \\
&& \dfrac{e^3}{(4\pi)^{\frac{D}{2}}} \Big\{ \Big( 2 (1-\xi) \Big[ \sint{2,2,2}{D+4} \left( p^\mu \slashed{k} - p \cdot k \gamma^\mu \right) \slashed{p}^\prime \nonumber \\
&& + \sint{2,2,2}{D+4} \left( p^{\prime \mu} \slashed{k} - p^\prime \cdot k \gamma^\mu  \right) \slashed{p} + 2 \sint{1,3,2}{D+4} \left( p^\mu \slashed{k} - p \cdot k \gamma^\mu \right) \slashed{p} \nonumber \\
&& + 2 \sint{3,1,2}{D+4} \left( p^{\prime \mu} \slashed{k} - p^\prime \cdot k \gamma^\mu  \right) \slashed{p}^\prime \Big] + \Big[ (4-D) \sint{1,1,1}{D} \nonumber \\
&& + (1-\xi)(\sint{1,1,1}{D}-2 \sint{1,1,2}{D+2}) \Big] \sigma^{\mu\alpha} k_\alpha  \Big) (\slashed{p}+m) \nonumber \\
&& + \Big[ 2 \sint{1,1,1}{D} + (D-6) \sint{1,2,1}{D+2} -(1-\xi)\Big( \sint{1,1,2}{D+2} \nonumber \\
&& + 2 p \cdot p^\prime \sint{2,2,2}{D+4} + 4 p^2 \sint{1,3,2}{D+4} - \sint{1,2,1}{D+2} \Big) \Big] \sigma^{\mu \alpha}k_\alpha \slashed{p} \nonumber \\
&& - \Big[ (6-D) \sint{2,1,1}{D+2} + (1-\xi)\Big( 2p^2 \sint{2,2,2}{D+4} \nonumber \\
&& + 4 \ppp \sint{3,1,2}{D+4} - \sint{2,1,1}{D+2} \Big) \Big] \sigma^{\mu \alpha }k_\alpha \slashed{p}^\prime \nonumber \\
&& + 2 \Big[ -2\sint{1,1,1}{D} + 2 \sint{1,2,1}{D+2} + (1-\xi)\Big( \sint{1,1,2}{D+2} \nonumber \\
&& + 2 \ppp \sint{2,2,2}{D+4}+4p^2 \sint{1,3,2}{D+4}-\sint{1,2,1}{D+2} \Big) \Big](p^\mu \slashed{k}-k \cdot p \gamma^\mu) \nonumber \\
&& + 2 \Big[ 2 \sint{2,1,1}{D+2} + (1-\xi)\Big( 2p^2 \sint{2,2,2}{D+4} + 4 \ppp \sint{3,1,2}{D+4} \nonumber \\
&& - \sint{2,1,1}{D+2} \Big) \Big](p^{\prime \mu}\slashed{k} -p^\prime \cdot k \gamma^\mu ) \Big\}. \label{VT}
\end{eqnarray}  
An alternative and more commonly adopted approach is to express the vertex $V^\mu$ by expanding it out in the Ball-Chiu basis, Eq.~\cite{vertex_any_gauge, Ball_chiu, vertex_QED_QCD}:
\begin{eqnarray}
V^\mu = \sum_{i=1}^4 \lambda_i L_i^\mu + \sum_{i=1}^8 \tau_i T_i^\mu, \label{Ball_Chiu}
\end{eqnarray}
where $L_i^\mu$ and $T_i^\mu$ are the longitudinal and the transverse basis vectors. $L_i^\mu$ are:
\begin{eqnarray}
L_1^\mu &=& \gamma^\mu,  \nonumber \\
L_2^\mu &=& (p+p^\prime)^\mu (\slashed{p} + \slashed{p}^\prime), \nonumber \\
L_3^\mu &=& (p+p^\prime)^\mu, \nonumber \\
L_4^\mu &=& \sigma^{\mu \alpha}(p+p^\prime)^\alpha, 
\end{eqnarray}
and
\begin{eqnarray}
T_1^\mu &=& p^\prime \cdot k p^\mu - p \cdot k p^{\prime \mu}, \nonumber \\
T_2^\mu &=& \left( p^\prime \cdot k p^\mu - p \cdot k p^{\prime \mu} \right) \left( \slashed{p}^\prime + \slashed{p} \right) \nonumber \\
T_3^\mu &=& k^2 \gamma^\mu - k^\mu \slashed{k}, \nonumber \\
T_4^\mu &=& -\left( p^\prime \cdot k p^\mu - p \cdot k p^{\prime \mu} \right) \sigma^{\alpha\beta} p^\prime_\alpha p_\beta, \nonumber \\
T_5^\mu &=& \sigma^{\mu\alpha} k_\alpha, \nonumber \\
T_6^\mu &=& \left( \pps - p^2 \right)\gamma^\mu -(p+p^\prime)^\mu \slashed{k}, \nonumber \\
T_7^\mu &=& -\dfrac{1}{2}\left( \pps - p^2 \right)\left[ \gamma^\mu (\slashed{p}^\prime + \slashed{p} ) + (p^\prime +p)^\mu \right] \nonumber \\
&& + (p^\prime +p)^\mu \sigma^{\alpha\beta} p^\prime_{\alpha} p_\beta, \nonumber \\
T_8^\mu &=& \gamma^\mu \sigma^{\alpha\beta}p^\prime_\alpha p_\beta + p^{\prime \mu} \slashed{p} -p^\mu \slashed{p}^\prime.  \label{Transverse_BC}
\end{eqnarray}
In order to provide a direct verification of our readily available and compact results, we can compare our findings with those reported in~\cite{vertex_QED_QCD} by projecting onto the above basis. Starting from the expressions for the longitudinal $V_L^\mu$ and transverse $V_T^\mu$  vertices given in Eqs.~\eqref{VL_2} and~\eqref{VT}, the coefficients $\lambda_i$ and $\tau_i$ of Eq.~\eqref{Ball_Chiu} can be identified in terms of scalar integrals as detailed in appendix~\ref{BC_coef}. These results
for the complete one-loop vertex in arbitrary gauge and dimensions are completely equivalent to the ones evaluated in~\cite{vertex_QED_QCD}.

\section{The Pauli form factor $F_2(k^2)$} \label{Form_factor}

The Pauli form factor $F_2(k^2)$ is defined through the on-shell matrix element of the vertex as follows:
\begin{eqnarray}
&& \bar{u}_{s'}(p') V^{\mu}(p',p) u_s(p) = \nonumber \\
&& e \bar{u}_{s'}(p') \Big[ F_1(k^2) \gamma^\mu  + \dfrac{1}{2m} F_2(k^2) \sigma^{\mu\nu} k_\nu \Big] u_s(p),  \label{Vertex_formfactors}
\end{eqnarray}
where $F_1(k^2)$ represents the Dirac form factor, and the Dirac spinors $u_s(p)$ and $\bar{u}_{s'}(p')$ satisfy
\begin{eqnarray}
\bar{u}_{s'}(p')(\slashed{p}^\prime+m) = (\slashed{p} + m) u_s(p) = 0. \label{spinors}
\end{eqnarray}
Using Eq.~\eqref{VL_1}, the matrix element of the longitudinal vertex $V_L^\mu$ reads as
\begin{eqnarray}
&& \hspace{-3mm} \bar{u}_{s'}(p') V^{\mu}_L(p',p) u_s(p) = \nonumber \\
 && \hspace{-3mm} \dfrac{e^3}{(4\pi)^{\frac{D}{2}}} \bar{u}_{s'}(p') \Bigg\{
2m(p+p^\prime)^\mu (2  \sint{2,1,1}{D+2} -\sint{1,1,1}{D}) \nonumber \\
&& \hspace{-3mm} + 2m(1-D+\xi) (p+p^\prime)^\mu \sint{2,1,1}{D+2}  \nonumber \\
&& \hspace{-3mm} + (2-D) \Big[ \sint{1,1,1}{D+2} \gamma^\mu - 2m (p+p^\prime)^\mu (2 \sint{3,1,1}{D+4}+\sint{2,2,1}{D+4})\Big] \nonumber \\
&& \hspace{-3mm} + (D-3+\xi) \sint{1,0,1}{D} \gamma^\mu  -  (p+p^\prime)^\mu \Big[ 2m(m^2-\ppp) \sint{2,2,2}{D+4} \nonumber \\
&& \hspace{-3mm} + 4m(m^2-\ppp) \sint{3,1,2}{D+4} - m \sint{1,1,2}{D+2} \Big] \Bigg\} u_s(p), \label{on_shell_VL_1} 
\end{eqnarray}  
where Eq.~\eqref{spinors} has been used together with the relation $\sint{\alpha,\beta, \gamma}{\tilde{D}} = \sint{\beta, \alpha, \gamma}{\tilde{D}}$, valid under on-shell condition $p^2 = \pps = -m^2$. In this section all the scalar integrals are taken on-shell, which means that the relation $p^2 = \pps = -m^2$ is satisfied.  
After applying Gordon identity,
\begin{eqnarray}
\hspace{-2.6mm} \bar{u}_{s'}(p') \left[ (p+p^\prime)^\mu + \sigma^{\mu\alpha}k_\alpha \right] u_s(p) \hspace{-0.5mm} = \hspace{-0.5mm} 2m \bar{u}_{s'}(p') \gamma^\mu u_s(p),
\end{eqnarray}
Eq.~\eqref{on_shell_VL_1} becomes
\begin{eqnarray}
&& \hspace{-9mm} \bar{u}_{s'}(p') V^{\mu}_L(p',p) u_s(p) = \dfrac{e^3}{(4\pi)^{\frac{D}{2}}} \times \nonumber \\
&& \hspace{-9mm}  \bar{u}_{s'}(p') \Big\{ 
\left[ (2-D) \sint{1,1,1}{D+2} + (D-3+\xi) \sint{1,0,1}{D} + f_{2\, L} \right] \gamma^\mu 
\nonumber \\
&& 
\hspace{6mm} - f_{2\,L} \; \sigma^{\mu\alpha} k_\alpha
\Big\} u_s(p), \label{on_shell_VL_2}
\end{eqnarray}
where
\begin{eqnarray}
f_{2\,L} &=& (1-\xi) m \Big[ \sint{1,1,2}{D+2} - 2 (m^2-\ppp)\sint{2,2,2}{D+4} \nonumber \\
&& -4 (m^2-\ppp) \sint{3,1,2}{D+4} -2\sint{2,1,1}{D+2} \Big] - 2 m \sint{1,1,1}{D} \nonumber \\
&&  + 4 m (D-2) \sint{3,1,1}{D+4} + 2(4-D) \sint{2,1,1}{D+2} \nonumber \\
&& - 2 m (2-D) \sint{2,2,1}{D+4}. 
\label{def_f2_L}
\end{eqnarray}
Similarly, we can evaluate the on-shell matrix element of the transverse vertex $V_T$ using Eq.~\eqref{VT}:
\begin{eqnarray}
&& \hspace{-1cm} \bar{u}_{s'}(p') V^{\mu}_T(p',p) u_s(p) = \dfrac{e^3}{(4\pi)^{\frac{D}{2}}} \times
\nonumber \\
&&  
\hspace{1.7cm} \bar{u}_{s'}(p') \left[
f_{1\, T} \gamma^\mu +f_{2\, T} \sigma^{\mu \alpha} k_\alpha \right] u_s(p), \label{on_shell_VT}
\end{eqnarray}
where
\begin{eqnarray}
f_{2\, T} &=& (1-\xi) m\Big[ \sint{1,1,2}{D+2} - 2 (m^2-\ppp) \sint{2,2,2}{D+4} \nonumber \\
&-& 4(m^2-\ppp) \sint{3,1,2}{D+4}-2\sint{2,1,1}{D+2} \Big] -2m \sint{1,1,1}{D} \nonumber \\
&+&  2(6-D)m \sint{2,1,1}{D+2}, \nonumber \\
f_{1\, T} &=& 4 k \cdot p \sint{1,1,1}{D} - 4 k \cdot p \sint{1,2,1}{D+2} + 2(4-D) k \cdot p^\prime \sint{2,1,1}{D+2} \nonumber \\
&-&  2 k \cdot p (1-\xi) \big( \sint{1,1,2}{D+2} + 2 \ppp \sint{2,2,2}{D+4} + 4 p^2 \sint{1,3,2}{D+4} \nonumber \\
&-&  \sint{1,2,1}{D+2}  \big).
\end{eqnarray}
According to Eq.~\eqref{Vertex_formfactors}, supplemented with the results of Eqs.~\eqref{on_shell_VL_2} and~\eqref{on_shell_VT}, the Pauli form factor $F_2(k^2)$ in any space-time dimension $D$ reads as,
\begin{eqnarray}
\hspace{-3mm} F_2(k^2) \hspace{-0.5mm} = \hspace{-0.5mm} \dfrac{4 e^2 m^2}{(4\pi)^{\frac{D}{2}}} \hspace{-0.5mm} \Big[ 2\sint{2,1,1}{D+2} + \hspace{-0.5mm} (2-D) \hspace{-0.8mm} \left( 2 \sint{3,1,1}{D+4} 
+ \sint{2,2,1}{D+4} \right) \hspace{-0.9mm} \Big],  \label{Pauli_FF}
\end{eqnarray}
which agrees with Eq.~(4.30) of Ref.~\cite{vertex_QED_QCD} after all the scalar integrals are represented as linear combination of on-shell master integrals.

It is important to notice the cancellation of the scalar integrals that are weighted by the factor $1-\xi$ in the Pauli form factor, implying its explicit gauge independence. 

\subsection{$F_2(k^2)$ in $D=4$ dimensions} \label{sec_F2_D4}

For $D=4$, Eq.~\eqref{Pauli_FF} for the Pauli form factor yields
\begin{eqnarray}
F_2(k^2) = -\dfrac{2 m^2 \alpha}{\pi} \left[ 2 \sint{3,1,1}{8} + \sint{2,2,1}{8} -\sint{2,1,1}{6} \right],
\end{eqnarray}
where $\alpha = e^2/(4\pi)$. The combination of scalar integrals above can be computed easily by a direct application of Feynman parametrization without the need to use the expressions given in the appendix \ref{s_int} to transform it in terms of master integrals. Thus it can be shown that,
\begin{eqnarray}
F_2 (k^2) = - \dfrac{2\alpha}{\pi} \int_0^1 dx \int_0^{1-x} dy \dfrac{x^2+xy-x}{x^2+y^2 + (2 + c)xy},
\end{eqnarray}
where $c={k^2}/{m^2}$.
After a convenient change of variables $y \xrightarrow{} x \left( {1}/{y} -1 \right) $ in the second integral, and interchanging the order of integration, the equation above becomes:
\begin{eqnarray}
F_2(k^2) &=& \dfrac{\alpha}{\pi} \int_0^1 dy\, \dfrac{y}{1 + c y (1-y)}. \label{F2_D4_1}
\end{eqnarray}
Since
\begin{equation}
\int_0^1 dy\, \dfrac{y}{1 + c y (1-y)} = \dfrac{1}{2} \int_0^1 dy\, \dfrac{1}{1 + c y (1-y)},
\end{equation}
Eq.~\ref{F2_D4_1} for the Pauli form factor in $D=4$ dimensions becomes
\begin{eqnarray}
F_2(k^2) = \dfrac{\alpha}{2\pi}  \int_0^1 \dfrac{dy}{1+cy(1-y)} ,  \label{F2_D4}
\end{eqnarray}
which agrees with the standard textbook result~\cite{Peskin}.   

\subsection{$F_2(k^2)$ in $D=3$ dimensions}

For $D=3$, Eq.~\eqref{Pauli_FF} for the Pauli form factor reads as
\begin{eqnarray}
    F_2(k^2) = \dfrac{4 e^2 m^2 }{(4 \pi)^{3/2}} \left( 2 \sint{2,1,1}{5} - 2\sint{3,1,1}{7} - \sint{2,2,1}{7} \right). 
\end{eqnarray}
In this case, the first scalar integral is infrared divergent. To regularize this divergence a fictitious photon mass $m_\gamma$ is introduced in the photon propagator. Thus, after Feynman parametrization, we have
\begin{eqnarray}
F_2(k^2) = \dfrac{e^2}{4\pi m} \int_0^1 dx \int_0^{1-x} dy \dfrac{x(2-x-y)}{ A^{3/2} }, \nonumber \\ \label{F_2_3D}
\end{eqnarray}
where $c=k^2/m^2$ as before, and
\begin{eqnarray}
A &=& (x+y)^2 + c x y + (1-x-y) \kappa^2, \nonumber \\
\kappa &=& m_\gamma/m. 
\end{eqnarray}
These integrals are easily evaluated when $c=0$, which defines the anomalous magnetic moment of the electron in 3-dimensions. After computing the integrals involved, Eq.~\eqref{F_2_3D} gives $F(k^2=0) \equiv F_2$:
\begin{eqnarray}
F_2 = \dfrac{e^2}{8 \pi m} \Big[ 3(\kappa -1) + \left( 2 - \dfrac{3}{2} \kappa \right) \ln \left( \dfrac{2+\kappa}{\kappa} \right) \Big], \label{AMM_D3}
\end{eqnarray}
which agrees with Eq.~(16) of Ref.~\cite{DasAnomalousMM}. It reaffirms that in 3-dimensions usual QED does not yield well-defined electron anomalous magnetic moment. However, it can be cured by adding a Chern-Simon term to the original Lagrangian as discussed in 
Ref.~\cite{DasAnomalousMM}.

\section{Summary and conclusions} \label{conclusions}

The second order formalism of QED is based on the Feynman rules depicted in Fig.~\ref{SecondOR}. It allows 
conceptual and clear decoupling
between scalar and spin degrees of freedom in an electromagnetic interaction of charged fermions. Any QED amplitude, constructed from these rules, is equivalent to the one obtained from the standard first order formalism. This equivalence is established by rewriting the product of the tree-level fermion propagator and the fermion-photon vertex as illustrated in Eq.~\eqref{sor_dec}, and applying identity~\eqref{C_prop}.

In this article, a hybrid perspective to study the Green functions in QED, such as the vertex $V^\mu$, is proposed. As the name suggests, it blends together the first and second order formalisms. It can be employed to obtain a more instructive Dirac and tensor structure of the Green functions. For the vertex, the combined analysis yields a natural separation between its longitudinal and transverse components. Before any attempt to perform the Feynman integrals is made, these components are given in Eqs.~\eqref{VL_l} and~\eqref{VT_l}.
There, the transversality condition for $V_T^\mu$ is clearly observed 
while the longitudinal part is demonstrated to satisfy the WTI.  

After the tensor reduction method is applied to the tensor Feynman integrals in Eq.~\eqref{VL_l} for $V_L^\mu$, and the contraction with the photon momenta $k_\mu$ is done, useful identities between scalar integrals with higher powers in the propagators and shifted dimension can be deduced. These identities are displayed in Eqs.~\eqref{iden1},~\eqref{iden2},~\eqref{iden3},~\eqref{iden4} and~\eqref{iden5}. The complicated scalar integrals that appear on its left hand side conspire in such a way as to obtain simple relations of master integrals with at most two propagators. Employing these identities one can demonstrate the WTI holds at one-loop order for the component $V_L^\mu$ of the vertex given in either Eqs.~\eqref{VL_l},~\eqref{VL_1} or~\eqref{VL_2}. 

The longitudinal and transverse separation that is obtained from the combination of the first and second order formalism also allows us to show explicitly, with minimum effort, the independence of the Pauli form factor $F_2$ from the gauge parameter $\xi$. In fact, the form factor in any dimensions has the compact expression obtained in Eq.~\eqref{Pauli_FF} in terms of scalar integrals. As shown in section~\ref{Form_factor}, in dimensions $D=4$ and $D=3$, this combination of the scalar integrals can be easily evaluated to recover known results in the literature.  

As a final remark the longitudinal and transverse decomposition obtained in this work are in agreement with the expressions obtained in Ref.~\cite{vertex_QED_QCD}.

\begin{acknowledgments}
A.B. acknowledges Coordinaci\'on de la
Investigaci\'on Cient\'ifica of the Universidad Michoacana
de San Nicol\'as de Hidalgo Grant No. 4.10., US
Department of Energy (DOE) under the Contract
No. DE-AC05-6OR23177 and the Fulbright-Garc\'ia
Robles Scholarship. V.M.B.G is grateful to Consejo Nacional de Ciencia
y Tecnología (Mexico) for support.
\end{acknowledgments}

\appendix


\section{The difference between the Ball-Chiu vertex  and $V_L^\mu$ at one-loop}

The results of Eq.~\eqref{VL_2} suggest the following vector structure for the expanded out longitudinal part $\Gamma_L^\mu$ of the full fermion-photon vertex $\Gamma^\mu$,
\begin{eqnarray}
\Gamma_L^\mu &=& a_1 p^\mu + a_2 p^{\prime \mu} + a_3 \gamma^\mu 
+ a_4 p^\mu \slashed{p} + a_5 p^\mu \slashed{p}^\prime \nonumber \\
&& + a_6 p^{\prime\mu} \slashed{p} + a_7 p^{\prime\mu} \slashed{p}^\prime,
\label{G_L}
\end{eqnarray}
where the scalar coefficients $a_i$ at one-loop are given by Eq.~\eqref{VL_2}. These coefficients are not entirely independent. Since, according to the WTI,
\begin{eqnarray}
k \cdot \Gamma_L^\mu &=& S^{-1}(p^\prime) - S^{-1}(p), \nonumber \\
S^{-1}(p) &=& F(p^2) \, \slashed{p} + G(p^2),
\end{eqnarray}
the following relations must be satisfied 
\begin{eqnarray}
a_1 k \cdot p + a_2 k \cdot p^\prime &=& G(p^{\prime 2}) - G(p^2), \nonumber \\
a_3 + a_5 k \cdot p + a_7 k \cdot p^\prime &=& F(p^{\prime 2}), \nonumber \\
-a_3 + a_4 k \cdot p + a_6 k \cdot p^\prime &=& - F(p^{2}). \label{a_iden}
\end{eqnarray}
On-shell, these relations are reduced to:
\begin{eqnarray}
a_1 &=& a_2, \nonumber \\
a_4 +a_5 &=& a_6 + a_7, \nonumber \\
a_3 + k \cdot p ( a_5 -a_7 ) &=& F(m^2). \label{a_iden_onshell}
\end{eqnarray}
The definition of $\Gamma_L^\mu$ in \eqref{G_L} differs from the Ball-Chiu vertex \cite{Ball_chiu},
\begin{eqnarray}
\Gamma_{BC}^\mu &=& \dfrac{G(p^{\prime 2})-G(p^2)}{p^{\prime 2}-p^2} (p+p^\prime)^\mu  \nonumber \\
&& + \dfrac{F(p^{\prime 2})+F(p^2)}{2} \gamma^\mu \nonumber \\
&& +  \dfrac{\slashed{p}^\prime + \slashed{p} }{2} (p+p^\prime)^\mu \dfrac{F(p^{\prime 2})-F(p^2)}{p^{\prime 2}-p^2}, 
\end{eqnarray}
by a transverse piece, which can be written in terms of the  coefficients $a_i$ as follows:
\begin{eqnarray}
 \delta \Gamma^\mu_L &=& \Gamma_L^\mu -  \Gamma_{BC}^\mu \nonumber  \\
 &=& \dfrac{a_1-a_2}{p^{\prime 2}-p^2} \; \tilde{T}_1^\mu  + \dfrac{a_4-a_5}{2} \; \tilde{T}_2^\mu  + \dfrac{a_6-a_7}{2} \; \tilde{T}_3^\mu  \nonumber \\
 && + \dfrac{a_4+a_5-a_6-a_7}{p^{\prime 2}-p^2} \; \tilde{T}_4^\mu, \label{delta_vertices}
\end{eqnarray}
where
\begin{eqnarray}
&& \hspace{-8mm} \tilde{T}_1^\mu =  k \cdot p^\prime p^\mu - k \cdot p p^{\prime \, \mu}, \nonumber \\
&& \hspace{-8mm} \tilde{T}_2^\mu = k \cdot p \gamma^\mu + p^\mu \slashed{p} - p^\mu \slashed{p}^\prime, \nonumber \\
&& \hspace{-8mm} \tilde{T}_3^\mu = k \cdot p^\prime \gamma^\mu + p^{\prime\mu} \slashed{p} - p^{\prime\mu} \slashed{p}^\prime, \nonumber \\
&& \hspace{-8mm} \tilde{T}_4^\mu = \Big[ k \cdot p^\prime \left(p^\mu \slashed{p} + p^\mu \slashed{p}^\prime \right) - k \cdot p \left(p^{\prime\mu} \slashed{p} + p^{\prime\mu} \slashed{p}^\prime \right) \Big]. 
\end{eqnarray}
In terms of the transverse Ball-Chiu basis defined earlier 
in~Eq.~(\ref{Transverse_BC}), these new vectors read as
\begin{eqnarray}
\tilde{T}_1^\mu &=& T_1^\mu, \nonumber \\
\tilde{T}_2^\mu &=& -\dfrac{1}{2} \left( T_3^\mu - T_6^\mu \right), \nonumber \\
\tilde{T}_3^\mu &=& \dfrac{1}{2} \left( T_3^\mu + T_6^\mu \right), \nonumber \\
\tilde{T}_4^\mu &=& T_2^\mu.
\end{eqnarray}
 Eq.~\eqref{delta_vertices} can be rewritten as follows in terms of scalar Feynman integrals at one-loop,
\begin{eqnarray}
 \delta \Gamma_L^{\mu\,1-L} &=& \delta V_L^\mu = \sum_{i=1}^4 \delta_i \tilde{T}_i^\mu, \label{delta_vertices_1L}
\end{eqnarray}
where 
\begin{eqnarray}
\delta_1 &=& \dfrac{2m(D-1+\xi)}{p^{\prime 2}-p^2} \left( \sint{1,2,1}{D+2} - \sint{2,1,1}{D+2} \right), \\
\delta_2 &=& \dfrac{1}{2} \Big[ (2-D) \big( \sint{1,1,1}{D} - 3 \sint{1,2,1}{D+2}+4\sint{1,3,1}{D+4} + \sint{2,1,1}{D+2} \nonumber \\ 
&&  - 2 \sint{2,2,1}{D+4} \big) + (1-\xi) \big( \sint{1,1,1}{D} - \sint{1,1,2}{D+2} - \sint{1,2,1}{D+2} \nonumber \\
&& - \sint{2,1,1}{D+4} + 4 m^2 \sint{1,3,2}{D+4} + 4 p\cdot p^\prime \sint{3,1,2}{D+4} \nonumber \\ 
&& - 2(m^2+p\cdot p^\prime)\sint{2,2,2}{D+4} \big)   \Big], \\
\delta_3 &=& \dfrac{1}{2} \Big[  (2-D) \big( \sint{1,1,1}{D} - \sint{1,2,1}{D+2} - \sint{2,1,1}{D+2} + 2\sint{2,2,1}{D+4} \nonumber \\
&& - 4\sint{3,1,1}{D+4} \big) + (1-\xi) \big( \sint{1,1,1}{D} - \sint{1,1,2}{D+2} + \sint{1,2,1}{D+2} \nonumber \\
&& - 3 \sint{2,1,1}{D+2} + 2(m^2-p\cdot p^\prime +2p^2 )\sint{2,2,2}{D+4} \nonumber \\
&& - 4(m^2-p \cdot p^\prime + p^2 ) \sint{3,1,2}{D+4} - 4 p^2 \sint{1,3,2}{D+4} \big) \Big], \\
\delta_4 &=& \dfrac{2}{p^{\prime 2}-p^2} \Big\{ (D-2) \big( \sint{1,2,1}{D+2}-2\sint{1,3,1}{D+4}-\sint{2,1,1}{D+2} \nonumber \\
&& +2\sint{3,1,1}{D+4} \big) + (\xi-1) \big[ \sint{1,2,1}{D+2} - \sint{2,1,1}{D+2} \nonumber \\
&& - 2 ( m^2 + p^2 )\big( \sint{1,3,2}{D+4} - \sint{3,1,2}{D+4} \big) \big] \Big\}.
\end{eqnarray}

\begin{widetext}

\section{Scalar integrals} \label{s_int}

In this section the Feynman scalar integrals that appear in the longitudinal and trasverse parts of the vertex, $V_L^\mu$ and $V_T^\mu$ respectively, in terms of master integrals are provided. Using the IBP technique \cite{IBP_alg, IBP_letter} and the dimensional recurrence relations \cite{Tarasov_DRR, Lee_DRR1, Lee_DRR2}, with the aid of the symbolic programming package LiteRed \cite{LiteRed1, LiteRed2}, the scalar integrals read as,

\begin{eqnarray}
\sint{1,2,1}{D+2} &=& \dfrac{1}{2\beta_1} \Big[ p \cdot p^\prime \sint{0,1,1}{D} -p^{\prime 2} \sint{1,0,1}{D} + (p^{\prime 2}-p\cdot p^\prime)\sint{1,1,0}{D} + ( p^{\prime 2} (m^2-k\cdot p)-m^2 p\cdot p^\prime )\sint{1,1,1}{D} \Big],  \\
\sint{2,1,1}{D+2} &=& \dfrac{1}{2\beta_1} \Big[ -p^2 \sint{0,1,1}{D} + p \cdot p^\prime \sint{1,0,1}{D} - k \cdot p \sint{1,1,0}{D} + (k\cdot p^\prime \, p^2 - k \cdot p \, m^2) \sint{1,1,1}{D}  \Big], \\
\sint{1,1,2}{D+2} &=& \dfrac{1}{2 \beta_1} \Big[ - k \cdot p \sint{0,1,1}{D} + k \cdot p^\prime \sint{1,0,1}{D} - k^2 \sint{1,1,0}{D} - k^2 (m^2-p \cdot p^\prime) \sint{1,1,1}{D} \Big], \\
\sint{1,1,1}{D+2} &=& \dfrac{1}{2(D-2) \beta_1 } \Big\{ (p^2 k \cdot p^\prime -m^2 k \cdot p)\sint{0,1,1}{D} + [p^{\prime 2}(m^2-k\cdot p)-m^2 p \cdot p^\prime]\sint{1,0,1}{D} \nonumber \\
&& -[k^2 m^4 + k^2 p^2 p^{\prime 2}-2m^2( p^{\prime 2} p\cdot p^\prime + p^2(p \cdot p^\prime - 2 p^{\prime 2}) )] \sint{1,1,1}{D} - k^2 (m^2-p \cdot p^\prime) \sint{1,1,0}{D} \Big\}, \\
\sint{2,2,1}{D+4} &=& \dfrac{1}{4(D-2)\beta_1^2} \left( \alpha^{(1)}_1 \sint{0,1,0}{D} + \alpha^{(1)}_2 \sint{0,1,1}{D} + \alpha^{(1)}_3 \sint{1,0,1}{D} + \alpha^{(1)}_4 \sint{1,1,0}{D} + \alpha^{(1)}_5 \sint{1,1,1}{D}  \right), \\
\sint{2,2,2}{D+4} &=& \dfrac{1}{4 \beta_1^2 \beta_2 } \left( \alpha^{(2)}_1 \sint{0,1,0}{D} + \alpha^{(2)}_2 \sint{0,1,1}{D} + \alpha^{(2)}_3 \sint{1,0,1}{D} + \alpha^{(2)}_4 \sint{1,1,0}{D} + \alpha^{(2)}_5 \sint{1,1,1}{D}  \right),
\end{eqnarray}
where
\begin{eqnarray}
\beta_1 &=& p^2 p^{\prime 2}-(p \cdot p^\prime)^2, \qquad
\beta_2 = (\pps + m^2)(p^2+m^2)k^2 - m^2 (\pps - p^2)^2,
\end{eqnarray}
and
\begin{eqnarray}
\alpha_1^{(1)} &=& (D-2)\beta_1, \nonumber \\
\alpha_2^{(1)} &=& -p^2 \left\{ (p \cdot p^\prime)^2 + \left[(D-2)p^2+(1-D)\ppp\right] p^{\prime 2} \right\} +m^2 \left\{ p^2\left[ (D-1) \ppp - (D-2)\pps \right] -(p \cdot p^\prime)^2  \right\}, \nonumber \\
\alpha_3^{(1)} &=& -m^2\left\{ (p \cdot p^\prime)^2 + (D-2)p^2 \pps -(D-1) \ppp \pps \right\} - \pps \left\{ (\ppp)^2 + p^2\left[ (1-D)\ppp +(D-2)\pps \right] \right\}, \nonumber \\
\alpha_4^{(1)} &=& (D-2)p^4 \pps +(\ppp)^2\left[ (D-4)\ppp + \pps \right] + p^2\left[ (\ppp)^2 - 3(D-2)\ppp \pps + (D-2) p^{\prime 4} \right] \nonumber \\
&& + m^2\left\{ p^2\left[ (1-D)\ppp + (D-2) \pps \right] + \ppp \left[ D \ppp +(1-D)\pps \right] \right\},
\nonumber \\
\alpha_5^{(1)} &=& m^2 \Big\{ D p^2 (\ppp)^2 + \left[ (D-2)p^4 -4(D-1) p^2 \ppp + D (\ppp)^2 \right] \pps + (D-2)p^2 p^{\prime 4} \Big\} \nonumber \\
&& + m^4\left\{ p^2 \left[ (1-D)\ppp +(D-2)\pps \right] + \ppp \left[ D \ppp +(1-D)\pps \right] \right\} \nonumber \\
&& +p^2 \pps \left\{ p^2\left[ (1-D)\ppp + (D-2)\pps \right] + \ppp \left[ D \ppp + (1-D)\pps \right] \right\}, \nonumber \\
\alpha_1^{(2)} &=& (D-2) \beta_1 \left[ \ppp \pps + p^2( \ppp - 2 \pps  ) - m^2 k^2 \right], \nonumber \\
\alpha_2^{(2)} &=& -p^2 \pps k^2 \left\{ (D-2) (\ppp)^2 + p^2\left[ (1-D) \ppp + \pps \right] \right\} + m^4 k^2 \left\{ -(\ppp)^2 + p^2 \left[ (D-1)\ppp -  (D-2) \pps \right] \right\} \nonumber \\
&& + m^2 \Big\{ (3-D)p^6 \pps + 2 (\ppp)^3 \pps + p^2 \ppp \left[ 2(D-2)(\ppp)^2 + (7-5D) \ppp \pps + 2(D-2) p^{\prime 4} \right] \nonumber \\
&& - p^4 \left[ (D+1)(\ppp)^2 + 2(1-2D) \ppp \pps + (1+D) p^{\prime 4} \right] \Big\}, \nonumber \\
\alpha_3^{(2)} &=& -p^2 \pps k^2 \left\{ (D-2)(\ppp)^2 + \pps \left[ p^2 + (1-D)\ppp \right] \right\} -m^4 k^2 \left\{ (\ppp)^2 + \left[ (D-2)p^2 + (1-D) \ppp \right] \pps \right\} \nonumber \\
&& + m^2 \Big\{ 2 p^2 (\ppp)^3  + \ppp \pps \left[ 2(D-2)p^4 + (7-5D)p^2 \ppp +2(D-2)(\ppp)^2 \right] \nonumber \\
&& -p^{\prime 4} \left[ (D+1)p^4 + 2(1-2D)p^2 \ppp + (1+D)(\ppp)^2 \right] -(D-3)p^2p^{\prime 6} \Big\},  \nonumber \\
\alpha_4^{(2)} &=& -m^4 k^2 \left\{ p^2 \left[ (D-1) \ppp - (D-2) \pps \right] + \ppp \left[ -D \ppp +(D-1) \pps \right] \right\} \nonumber \\
&& + p^2 \pps k^2 \left\{ p^2 \left[ (1-D)\ppp + (D-2)\pps \right] + \ppp \left[ D \ppp + (1-D)\pps \right] \right\} \nonumber \\
&& + m^2 \Big\{ (D-3) p^6 \pps + (\ppp)^2 \pps \left[ -2D \ppp + (1+D) \pps \right] \nonumber \\
&&+ p^4 \left[ (D+1)(\ppp)^2 + 2(4-3D)\ppp \pps +2(D-1) p^{\prime 4} \right] \nonumber \\
&&+ p^2 \left[ -2D(\ppp)^3 + 10(D-1)(\ppp)^2 \pps + 2(4-3D)\ppp p^{\prime 4} + (D-3) p^{\prime 6} \right] \Big\}, \nonumber \\
\alpha_5^{(2)} &=& p^2 \pps k^2 \left\{ (D-2)(p^2-\ppp)(\ppp)^2 + \pps \left[ p^4 - D p^2 \ppp + (D-2)(\ppp)^2 \right] + p^2 p^{\prime 4} \right\} \nonumber \\
&& +m^4 \Big\{ (\ppp)^2 \left[ 2D p^4 +(6-5D)p^2 \ppp + 2 (D-2)(\ppp)^2 \right] \nonumber \\
&& + \pps \left[ (D-3)p^6 + (6-7D)p^4 \ppp +2(7D-8)p^2(\ppp)^2 + (6-5D)(\ppp)^3 \right] \nonumber \\
&& p^{\prime 4}\left[ 2(1+D)p^4 + (6-7D)p^2 \ppp +2D(\ppp)^2 \right] + (D-3)p^2 p^{\prime 6} \Big\}
\nonumber \\
&& + m^2 \Big\{ -p^2 (\ppp)^3 \left[ D(p^2-2\ppp)+4\ppp \right] \nonumber \\
&& + \pps \ppp \left[ (3-2D)p^6 + (9D-10)p^4 \ppp -12(D-2)p^2(\ppp)^2 + 2(D-2)(\ppp)^3 \right] \nonumber \\
&& + p^{\prime 4}\left[ (D+2)p^6 -6(D+1)p^4 \ppp + (9D-10)p^2 (\ppp)^2 - D (\ppp)^3 \right] \nonumber \\
&& + p^2 p^{\prime 6}\left[ (D+2)p^2 + (3-2D)\ppp \right] \Big\} \nonumber \\
&& -m^6 k^2 \left\{ p^2 \left[(D-1)\ppp - (D-2) \pps \right] + \ppp \left[ -D \ppp + (D-1) \pps \right] \right\}.
\end{eqnarray}
According to the identities \eqref{iden2}, \eqref{iden3}, \eqref{iden4} and \eqref{iden5}, the remaining scalar integrals, $\sint{3,1,1}{D+4}$, $\sint{1,3,1}{D+4}$, $\sint{3,1,2}{D+4}$ and $\sint{1,3,2}{D+4}$, can be expressed as a linear combination of the others as,
\begin{eqnarray}
\sint{3,1,1}{D+4} &=& \dfrac{1}{4 k \cdot p^\prime} \left[ \dfrac{1}{2 \pps} \sint{0,1,0}{D}  + \dfrac{\pps+m^2}{2 \pps} \sint{1,0,1}{D} - 2 k \cdot p \sint{2,2,1}{D+4} - \sint{1,1,1}{D+2} -(p^2 - \pps) \sint{2,1,1}{D+2} \right],  \nonumber \\
\sint{1,3,1}{D+4} &=& \dfrac{1}{4 k \cdot p} \left[ \sint{1,1,1}{D+2} -(p^2-\pps)\sint{1,2,1}{D+2} - 2 k \cdot p^\prime \sint{2,2,1}{D+4} - \dfrac{1}{2p^2} \sint{0,1,0}{D} - \dfrac{p^2+m^2}{p^2} \sint{0,1,1}{D} \right], \nonumber
\end{eqnarray}
\begin{eqnarray}
\sint{3,1,2}{D+4} &=& \dfrac{1}{4 \pps (p^2+m^2)-4 \ppp (\pps+m^2)} \Bigg[ \dfrac{1}{2 p^{\prime 2}} \left\{ \left[ (D-4) p^{\prime 2} - (D-2) m^2\right] \sint{1,0,1}{D}-(D-2)\sint{0,1,0}{D} \right\}
\nonumber \\
&& - (p^{\prime 2}-p^2) \sint{2,1,1}{D+2} -(p^2+m^2)\left( \sint{1,1,2}{D+2} + 2 p \cdot p^\prime \sint{2,2,2}{D+4} \right) +2 p^2(p^{\prime 2}+m^2) \sint{2,2,2}{D+4}  \Bigg], \nonumber \\
\sint{1,3,2}{D+4} &=& \dfrac{1}{4 \ppp (p^2+m^2)-4p^2(\pps +m^2)} \Bigg[ \dfrac{1}{2 p^2} \left\{ \left[ (D-4)p^2 - (D-2) m^2 \right] \sint{0,1,1}{D} - (D-2) \sint{0,1,0}{D} \right\} \nonumber \\
&& - (p^{\prime 2}-p^2) \sint{1,2,1}{D+2} - 2 p^{\prime 2} (p^2+m^2) \sint{2,2,2}{D+4} + (p^{\prime 2}+m^2) \left( \sint{1,1,2}{D+2} + 2 p \cdot p^\prime \sint{2,2,2}{D+4}  \right) \Bigg].
\end{eqnarray}

\section{Ball-Chiu coefficients} \label{BC_coef}

Using the results given in Eqs.~\eqref{VL_2} and~\eqref{VT}, the coefficients $\lambda_i$ in Eq.~\eqref{Ball_Chiu} reads as
\begin{eqnarray}
\lambda_1 &=& \dfrac{1}{2} (D-2) \Big[ 2 \sint{1,0,1}{D} + (\pps-p^2)\sint{1,1,1}{D} - 2 \sint{1,1,1}{D+2} + (3p^2 - 2 \ppp -\pps)\sint{1,2,1}{D+2} + (4 \ppp -4p^2) \sint{1,3,1}{D+4} \nonumber \\
&& + (2 \ppp-p^2-\pps)\sint{2,1,1}{D+2} + (2p^2 - 4 \ppp + 2 \pps) \sint{2,2,1}{D+4} + 4(\ppp - \pps) \sint{3,1,1}{D+4} \Big] \nonumber \\
&& + \dfrac{1}{2}(1-\xi)\Big\{ - 2 \sint{1,0,1}{D} + (p^2-\pps) \sint{1,1,1}{D} + (2m^2+p^2+\pps) \sint{1,1,2}{D+2} + (2\ppp-p^2-\pps) \sint{1,2,1}{D+2} \nonumber \\
&& + 4(m^2 p^2 - m^2 \ppp - p^2 \ppp + p^2 \pps) \sint{1,3,2}{D+4} + (3\pps -p^2 - 2\ppp ) \sint{2,1,1}{D+2}
\nonumber \\
&& + 2(2m^2\ppp -m^2p^2 + p^2 \ppp - m^2 \pps - 2 p^2 \pps + \ppp \pps ) \sint{2,2,2}{D+4} \nonumber \\
&& + 4( \pps [m^2+p^2-\ppp] -m^2 \ppp )\sint{3,1,2}{D+4} \Big\}, \nonumber \\
\lambda_2 &=& \dfrac{1}{2(p^2-\pps)} \Bigg\{ (D-2) \Big[ (\pps - p^2) \sint{1,1,1}{D} + (3p^2 - 2 \ppp -\pps)\sint{1,2,1}{D+2} + 4(\ppp - p^2) \sint{1,3,1}{D+4} \nonumber \\
&& + (p^2 + 2 \ppp - 3 \pps ) \sint{2,1,1}{D+2} + (2\pps-2p^2) \sint{2,2,1}{D+4} + 4(\pps -\ppp)\sint{3,1,1}{D+4} \Big] \nonumber \\
&& (1-\xi) \Big[ (p^2-\pps) \sint{1,1,1}{D} + (\pps -p^2) \sint{1,1,2}{D+2} + (2\ppp -p^2 - \pps) \sint{1,2,1}{D+2} \nonumber \\
&& + 4(m^2 p^2 - m^2 \ppp - p^2 \ppp + p^2 \pps ) \sint{1,3,2}{D+4} + (p^2 - 2\ppp +\pps) \sint{2,1,1}{D+2} \nonumber \\
&& + 2 ( m^2p^2 - p^2 \ppp - m^2 \pps + \ppp \pps) \sint{2,2,2}{D+4} + 4 \Big(m^2 \ppp -\pps(m^2+p^2-\ppp) \Big)\sint{3,1,2}{D+4} \Big] \Bigg\}, \nonumber \\
\lambda_3 &=& \dfrac{m(D-1+\xi)}{\pps - p^2} \left[ (p^2 - \pps) \sint{1,1,1}{D} + 2(\ppp - p^2) \sint{1,2,1}{D+2} + 2(\pps - \ppp) \sint{2,1,1}{D+2} \right], \nonumber \\
\lambda_4 &=& 0,
\end{eqnarray}
while the coefficients $\tau_i$ correspond to

\begin{eqnarray}
\tau_1 &=& \dfrac{2m(D-1+\xi)}{\pps -p^2} \left( \sint{1,2,1}{D+2} - \sint{2,1,1}{D+2} \right), \nonumber \\
\tau_2 &=& \dfrac{1}{p^2 - \pps} \Big\{ (2-D) \left( \sint{1,2,1}{D+2} -2 \sint{1,3,1}{D+4} - \sint{2,1,1}{D+2} + 2 \sint{3,1,1}{D+4} \right) \nonumber \\
&& + (1-\xi) \left[ \sint{1,2,1}{D+2} - 2(m^2+p^2) \sint{1,3,2}{D+4} - \sint{2,1,1}{D+2} + (\pps - p^2) \sint{2,2,2}{D+4} + 2 (m^2+\pps)\sint{3,1,2}{D+4}  \right] \Big\}, \nonumber \\
\tau_3 &=& \dfrac{1}{2} \Big\{ (2-D) \left[ \sint{1,1,1}{D} - 2 \left( \sint{1,3,1}{D+4} - \sint{2,2,1}{D+4} + \sint{3,1,1}{D+4} \right) \right] + (1-\xi) \Big[ \sint{1,1,1}{D} - \sint{1,1,2}{D+2} \nonumber \\
&& - 2(m^2+p^2) \sint{1,3,2}{D+4} + (2m^2+p^2 + \pps) \sint{2,2,2}{D+4} - 2(m^2+\pps) \sint{3,1,2}{D+4} \Big] \Big\}, \nonumber 
\end{eqnarray}
\begin{eqnarray}
\tau_4 &=& \dfrac{4m(1-\xi)}{p^2-\pps} \left( \sint{1,3,2}{D+4} - \sint{3,1,2}{D+4} \right), \nonumber \\
\tau_5 &=& (4-D) m \sint{1,1,1}{D} + m (1-\xi) \Big[ \sint{1,1,1}{D} - 2 \sint{1,1,2}{D+2} + 2(\ppp -p^2) \sint{1,3,2}{D+4} \nonumber \\
&& + (p^2-2\ppp + \pps) \sint{2,2,2}{D+4} + 2(\ppp - \pps) \sint{3,1,2}{D+4} \Big], \nonumber \\
\tau_6 &=& \dfrac{1}{2} \Big\{ (D-2) \left( \sint{1,2,1}{D+2} -2 \sint{1,3,1}{D+4} - \sint{2,1,1}{D+2} + 2 \sint{3,1,1}{D+4} \right) + (1-\xi) \Big[ \sint{1,2,1}{D+2} + 2 (m^2-p^2) \sint{1,3,2}{D+4} \nonumber \\
&& - \sint{2,1,1}{D+2} + (p^2 -\pps) \sint{2,2,2}{D+4} + 2(\pps - m^2) \sint{3,1,2}{D+4} \Big] \Big\}, \nonumber \\
\tau_7 &=& \dfrac{2m(1-\xi)}{\pps - p^2} \left[ 2(\ppp-p^2) \sint{1,3,2}{D+4} + (\pps -p^2) \sint{2,2,2}{D+4} + 2 (\pps - \ppp) \sint{3,1,2}{D+4} \right], \nonumber \\
\tau_8 &=& (6-D) \left( \sint{1,1,1}{D} - \sint{1,2,1}{D+2} -\sint{2,1,1}{D+2} \right) + (1-\xi) \Big( \sint{1,1,1}{D} -3 \sint{1,1,2}{D+2} + \sint{1,2,1}{D+2} - 4p^2 \sint{1,3,2}{D+4} \nonumber \\
&& + \sint{2,1,1}{D+2} - 4 \ppp \sint{2,2,2}{D+4} - 4 \pps \sint{3,1,2}{D+4} \Big).
\end{eqnarray}

All these coefficients agree with the results presented in \cite{vertex_QED_QCD} after changing to its conventions, replacing $1-\xi$ by $\xi$, and using the expressions of the appendix \ref{s_int} to rewrite the coefficients in terms of master integrals.

\end{widetext}



\bibliographystyle{apsrev4-1}

\bibliography{references.bib}

\end{document}